\documentclass[compsoc, conference, a4paper, 10pt, times]{IEEEtran}

\usepackage{cite}
\usepackage{amsmath,amssymb,amsfonts}
\usepackage{algorithmic}
\usepackage{graphicx}
\usepackage{textcomp}
\usepackage{xcolor}
\usepackage{booktabs}
\usepackage[hidelinks]{hyperref}
\usepackage{amssymb}
\usepackage{wasysym}

\begin{document}

\title{SoK: Analysis of Root Causes and Defense Strategies for Attacks on Microarchitectural Optimizations}


\author{
\IEEEauthorblockN{Nadja Ramh{\"o}j Holtryd, Madhavan Manivannan and Per Stenstr{\"o}m}
\IEEEauthorblockA{
    \textit{Department of Computer Science and Engineering}\\ 
    \textit{Chalmers University of Technology}\\
    \textit{G{\"o}teborg, Sweden}\\
    \textit{Email: \{holtryd, madhavan, per.stenstrom\}@chalmers.se}\\
    }
}

\maketitle

\begin{abstract}
Microarchitectural optimizations are expected to play a crucial role in ensuring performance scalability. 
However, recent attacks have demonstrated that microarchitectural optimizations, which were assumed to be secure, can be exploited. Moreover, new attacks surface at a rapid pace limiting the scope of existing defenses. 
These developments prompt the need to review  microarchitectural optimizations with an emphasis on security, understand the attack landscape and the potential defense strategies. 

We analyze timing-based side-channel attacks targeting a diverse set of microarchitectural optimizations. We provide a framework for analysing non-transient and transient attacks, which highlights the similarities. 
We identify the four 
root causes of timing-based side-channel attacks: \textit{determinism}, \textit{sharing}, \textit{access violation} and \textit{information flow}, through our systematic analysis. 
\textcolor{black}{Our key insight is that a subset (or all) of the root causes are exploited by attacks and eliminating any of the exploited root causes, in any attack step, is enough to provide protection.} 
Leveraging our framework, we systematize existing defenses and show that they target these root causes in the different \textcolor{black}{attack steps.} 
\end{abstract}



\vspace{-.2em}
\section{Introduction}
\vspace{-.2em}
\label{intro}
Computer architecture is facing a security crisis~\cite{Hennessy,hill}. Recent attacks ~\cite{spectre,Meltdown,rowhammer_14, RIDL_19, foreshadow_18, AEPICLEAK,ZombieLoad_mds_19} have demonstrated that microarchitectural optimizations, which were assumed to be fundamentally secure for a long time, leak information which can be exploited 
to steal secrets.
Furthermore, efficient attacks continuously emerge targeting defenses, thereby limiting their effectiveness or even rendering the defenses moot altogether~\cite{speculative_probing_20,Speculative_interference_attacks_21,SpectreRewind_20,vila19,CEASAR-S,brutus,adv_profiling,casa20,Purnal21,song21,BranchHistoryInjection22,RETBLEED_22}. Simultaneously, with the slowing down of Moore's Law, microarchitectural optimizations are expected to play an increasingly important role in ensuring performance scalability. Consequently, there is a strong need to be able to leverage microarchitectural optimizations without compromising 
security. 


Microarchitectural optimizations, widely implemented in commercial processors, like branch predictors~\cite{predicitingkeys07,SimplePredictionAnalysis,jumpOverASLR16,spectre,BranchScope18}, caches~\cite{flush_flush, flush_reload, prime_probe} and prefetchers~\cite{amdAttacksSpaceTime,chen2021leaking,fetchingTale,unvelining,prefetchSCA,augury22} among others, are prone to attacks. A recent paper~\cite{Pandora} demonstrates that several optimizations proposed in literature, but not known to be commercially implemented as yet, such as value prediction~\cite{VP_96}, also are vulnerable. 
This underscores the importance of conducting a thorough review of microarchitectural optimizations with an emphasis on security.

\textcolor{black}{Prior works have started the important task of analyzing attacks and defenses for different microarchitectural optimizations~\cite{Szefer2019,SysEvauationTransient19,EvolutionDefencesTransient20,sok_hw_defences_21,survey_microarch_21,sok_spectre_sw_21,ruby2021,transient_ecex_attacks_22,Ge2018}. However, most of the works focus only on transient 
attacks and defenses~\cite{SysEvauationTransient19,EvolutionDefencesTransient20,sok_hw_defences_21,transient_ecex_attacks_22,ruby2021}, SW-based defenses~\cite{sok_spectre_sw_21} or cover a limited set of non-transient 
attacks~\cite{Szefer2019,Ge2018}. Pandora~\cite{Pandora} considers a broader set of non-transient 
microarchitectural optimizations and provides \emph{microarchitectural leakage descriptors} (MLDs) which quantify the information leakage. The MLDs show if a specific optimization can leak and how much information is leaked (1-bit or a few bits). 
Unfortunately, this information falls short on providing a systematic analysis of the similarities across different 
microarchitectural optimizations and the underlying root causes which make them vulnerable to attacks. Such an analysis  
can also help with the categorization of existing defense strategies and with the potential identification of attacks and defenses.}

Our goal, in this paper, is to 
perform a systematic analysis to highlight the common root causes which make microarchitectural optimizations vulnerable to exploits \textcolor{black}{that reveal secrets}. 
In order to enable analysis of a diverse set of microarchitectural optimizations, we present an abstract model of the architecture and the microarchitectural state transitions involved in an attack. 
Using this model as a framework, we analyze several timing-based side-channel 
attacks available in the literature on an extensive set of microarchitectural optimizations: cache, prefetching, branch prediction, computational simplification, speculative execution and value prediction. We also analyse additional microarchitectural optimizations like cache compression, pipeline compression, register-file compression, silent stores and computation reuse but omit them from the discussion due to space constraints. 

\textcolor{black}{Our analysis reveals four root causes which are exploited 
in order to succeed with attacks targeting the diverse set of microarchitectural optimizations covered.} 
The root causes are \textbf{determinism}, \textbf{sharing}, \textbf{access violation} and \textbf{information flow}. 
Here, determinism causes microarchitectural optimizations to be triggered in the same way under the same pre-conditions, leading to  predictable microarchitectural state transitions and timing variations.
Sharing of microarchitectural state, which is accessible to both the adversary and the victim, enables the creation of a side-channel.
Access violation enables access to a secret outside of the intended protection domain. Finally, information flow refers to exchange of information through microarchitectural state.
We note that a subset of these 
root causes have been identified individually in the context of specific attacks~\cite{RandomFillCache,practicalMitig,Szefer2019,sok_hw_defences_21,AEPICLEAK}. However, in our analysis we show that a subset of, or all, the root causes are common across attacks on a broad set of microarchitectural optimizations.

We show that the proposed defenses that focus on the vulnerabilities in different microarchitectural optimizations can be classified as targeting one or more of the identified 
root causes. We observe that similar defenses can be / are applied across different microarchitectural optimizations, with the same root cause vulnerability
. \textcolor{black}{For instance, partitioning can thwart attacks using the cache~\cite{DAWG_MICRO,CATalyst}, SMT~\cite{SecSMT} and branch prediction~\cite{BRB,HyBP_22}, by affecting sharing, information flow and determinism.} In addition, the defenses can also be applied to address the applicable root causes 
in the different steps of the attack. \textcolor{black}{Eliminating any of the root causes, exploited by an attack, in any of the attack steps can protect against the attack.}
We also discuss \textcolor{black}{potential attacks and defenses} for vulnerable microarchitectural optimizations.

Overall, our analysis demonstrates the versatility of the framework to capture a diverse set of attacks and defense strategies for different microarchitectural optimizations. We expect that our framework can be easily extended to study microarchitectural optimizations we do not explicitly cover in this paper.
We also believe that it can assist computer architects in understanding the landscape of attacks on a broad range of microarchitectural optimizations, categorizing existing defense strategies proposed to thwart such attacks, and in designing secure microarchitectural optimizations.

In summary, we make the following contributions:
\begin{itemize}
  \item We identify four root causes: \textbf{determinism}, \textbf{sharing}, \textbf{access violation} and \textbf{information flow}, that enable timing-based side-channel attacks on a wide range of microarchitectural optimizations.
  \item We provide a framework and 
  analyze both transient and non-transient execution attacks on a broad range of microarchitectural optimizations, highlighting similarities and differences. 
  \item We analyze available defenses using our framework and make a classification based on the root causes they address. 
  Based on the analysis, we discuss potential attack and defense possibilities for microarchitectural optimizations. 
\end{itemize}

The paper is structured as follows. Section \ref{background} presents our framework and defines the root causes. Section \ref{root_causes} and \ref{systematic_defenses} use the framework to systematize the attacks and defenses, respectively. 
Section \ref{discussion} discuss general observations before we conclude in Section \ref{conclusions}.  

\section{Systematization Framework}
\label{background}


We first present an abstract architecture model  and outline the steps for carrying out attacks based on the model. We then use this model as a framework to identify the root causes that enable attacks. Finally, we present an actual attack in the context of this framework.


\subsection{Abstract model and side-channel attack}
\label{optimization}
The architecture model is represented as a finite state machine (FSM) where the \textit{architectural state (AS)}, comprising SW-visible registers and memory, is the externally visible interface, that is accessible to a program. An FSM transition is caused when instruction execution 
leads to a change in the $AS$. 

The microarchitecture represents an implementation of the FSM specification and typically comprises several microarchitectural optimizations, denoted  
\emph{\{O\textsubscript{1},O\textsubscript{2}, ... ,O\textsubscript{n}\}}, to enable an efficient implementation.  A microarchitectural optimization uses a set of microarchitectural resources, denoted 
\emph{R\ =\  \{R\textsubscript{1},R\textsubscript{2}, ... ,R\textsubscript{m}\}}, to implement the intended functionality. 
This model permits resources to be shared across different optimizations. We define \textit{microarchitectural state (MS)} as a snapshot of the state of all the $m$ microarchitectural resources in the system  at time instance $t$, denoted \emph{MS = \{state(R\textsubscript{1}), state(R\textsubscript{2}), ... , state(R\textsubscript{m})\}\textsubscript{t}}. 

It is important to note that while the change in $AS$ 
caused by an FSM transition remains the same across different implementations of a given FSM specification, the change in \emph{MS} varies depending on the optimizations triggered, resources used and the implementation.
Even when considering a specific implementation, there is a one-to-many mapping relationship between $AS$ and \emph{MS}; i.e., a single $AS$ can have several equivalent \emph{MS}. Furthermore, the time it takes for an implementation to make a transition between different \emph{MS} (caused by an action) may vary and this property is typically exploited by attacks. \textcolor{black}{As an example, executing a load instruction will cause a change in the $AS$ while the insertion of a corresponding line in the cache hierarchy as a consequence of executing the load instruction will cause a change in the state of the cache(s) which is a microarchitectural resource.}

\begin{figure}[b!]
\centering
\vspace{-1.em}
\includegraphics[width=0.47\textwidth]{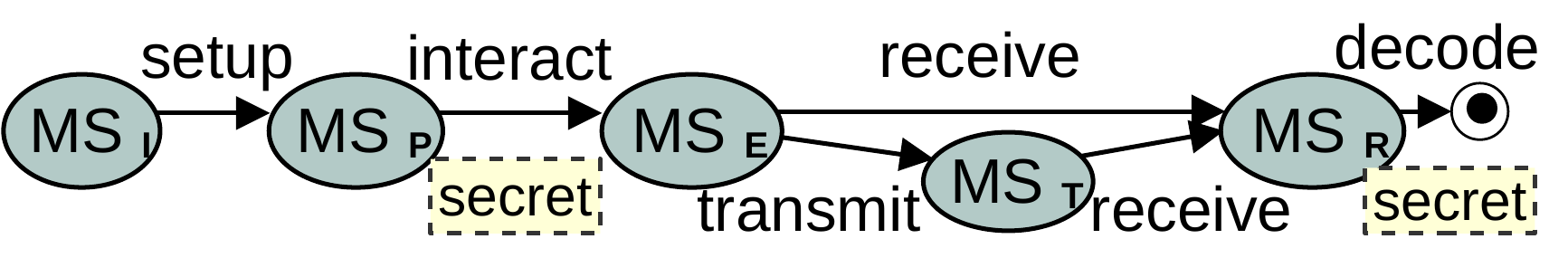}
\vspace{-1em}
\caption{\emph{MS} transitions in different steps of an attack.}
\label{fig:side_channel_attack}
\end{figure}

We next consider an abstract model of an attack that shows the different steps involved while leveraging \emph{MS} as the side-channel to communicate the secret from a victim to an adversary. In our model, we define a step as a tuple of current state and action which leads to a new state, \emph{\{MS\textsubscript{current}, action}\}$\rightarrow$\emph{MS\textsubscript{next}}. Figure~\ref{fig:side_channel_attack} shows the different steps listed in this model and is based on the attacks proposed in literature~\cite{Szefer2019,SysEvauationTransient19,EvolutionDefencesTransient20,sok_hw_defences_21,survey_microarch_21,sok_spectre_sw_21,ruby2021,transient_ecex_attacks_22}. We assume \emph{MS} is in the initial state (\emph{MS\textsubscript{I}}) before any of the steps in the attack are carried out. When the  \emph{setup} step is performed \emph{MS\textsubscript{I}} makes a transition to the primed state (\emph{MS\textsubscript{P}}).
The  \emph{setup} step ensures that the necessary preconditions are in place to encode the secret into \emph{MS\textsubscript{P}} in the next step of the attack.  
When the \emph{interact} step is performed the secret is accessed and is encoded in the microarchitectural state (\emph{MS\textsubscript{E}}). The secret is encoded  specifically through the state of one or more microarchitectural resources. If the secret is encoded through  a microarchitecture resource state, that is accessible to both the victim and the adversary, it can potentially be used as a side-channel to communicate the secret. 

Attacks optionally utilize the \emph{transmit} step in case the encoded microarchitectural resource state is not accessible by the adversary or the specific \emph{MS} based side-channel is noisy (i.e. the channel is prone to high error rate and has low channel bandwidth). When the transmit action is performed the secret is usually re-encoded through the state of a different shared microarchitectural resources (\emph{MS\textsubscript{T}}) 
which can address the aforementioned transmission limitations. 
When the \emph{receive} step is performed, the adversary accesses the microarchitectural state of the specific resource(s) and observes timing variations based on the encoded secret while the state transitions to \emph{MS\textsubscript{R}}. 
Finally, in the decode step, the timing variations observed 
are used as the basis to infer the secret.

The steps outlined above that cause \emph{MS} transitions and secret information to be leaked can be performed by the adversary, the victim or both, depending on the type of  attack. In the abstract model it is required that the state of at least one microarchitectural resource is shared between the victim and an adversary to enable information flow and consequently communicate the secret. The microarchitectural resources that are shared between the victim and the adversary are specific to the implementation and the threat model (see Section \ref{sec:threat_model} for details).

\textcolor{black}{Prior works define attack steps differently which leads to fewer/more steps. 
For example, Xiong et al. \cite{survey_microarch_21} defines three attack steps while Hu et al.\cite{sok_hw_defences_21} use six attack steps. In contrast, our attack model include five steps where each step consists of action(s) performed by adversary and/or victim on microarchitectural resource(s) \emph{MS} which leads to a new \emph{MS}. 
Note that the difference between the \emph{interact} and \emph{transmit} step is that the former accesses and encodes the secret on \emph{MS} while the latter re-encodes the secret on shared \emph{MS}.}


\textcolor{black}{We exemplify 
the abstract model 
by describing the steps in the well-known flush+reload~\cite{flush_reload} attack. This attack uses a single shared microarchitectural resource, a shared cache (\emph{SC}). 
The goal of the attack is to infer the secret which is revealed through the victim's cache accesses because of data-dependent control flow. A prerequisite for the attack is that the cache lines of interest are mapped to a shared page that is accessible by the adversary as well as the victim. During the  \emph{setup} step, the adversary uses the  \emph{clflush} instruction to evict the lines belonging to the shared pages from the cache. The fact that the pages are shared allows the adversary to evict data that is accessed by a victim. The state of the cache after this step is \emph{MS\textsubscript{P}\{R\textsubscript{SC}[[target]=null]\}}. During the \emph{interact} step the victim executes and interacts with the secret which is encoded in the \emph{SC} state by the presence/absence of the specific target cache line(s). The cache state changes to \emph{MS\textsubscript{E}\{R\textsubscript{SC}[[target]=A]\}}. Since the cache is shared the adversary can detect the state change that has occurred as a result of the interaction. The adversary, during the \emph{receive} step, accesses the cache line(s) (target) and measures the time. Through timing the adversary deduces which line(s) the victim has inserted and thereby infers the secret.}

\vspace{-.2em}
\subsection{Root causes}
\vspace{-.2em}
We 
define the 
root causes of an attack in the context of the abstract model and exemplify 
\textcolor{black}{with} an actual attack. 

\subsubsection{Determinism}
\label{sec:determinism}
We define determinism as the characteristic 
of a microarchitectural optimization whereby microarchitectural resource(s) used by an optimization, under the same pre-conditions, is/are triggered in the same manner and cause a predictable microarchitectural state transition and timing variation.
In other words, determinism causes an expected \emph{MS} transition and timing variation upon an action by the adversary and/or the victim. In the abstract model of the attack, determinism enables the adversary to control \emph{MS} transitions from \emph{MS\textsubscript{I}} through to \emph{MS\textsubscript{R}} across multiple steps.

\subsubsection{Sharing}
\label{sec:sharing}

We define sharing as the characteristic 
of a microarchitectural optimization whereby the state of the microarchitectural resource(s) used by an optimization is/are shared between a victim and an adversary. 
In the abstract model sharing allows for the creation of a side-channel between the victim and the adversary's protection domain through \emph{MS}. 

\subsubsection{Access violation}
\label{sec:access_violation}
We define access violation as the characteristic 
of a microarchitectural optimization whereby one/many microarchitectural resource(s) permit(s) access of secret data which is outside the protection domain of the program \textcolor{black}{on the microarchitectural level}. This consequently enables information to flow outside the intended protection domain and occurs either in the \emph{interact} or the \emph{receive} step of the attack 
which causes secret information to be encoded into \emph{MS}.


\subsubsection{Information flow}
\label{sec:information_flow}
We define information flow as the characteristic 
of a microarchitectural optimization to exchange information through the state of one or many 
microarchitectural resource(s). Information flow enables the adversary to infer the secret 
by observing the state change of microarchitectural resource(s). 



\textcolor{black}{
The flush+reload attack, discussed earlier, exploits determinism, sharing and information flow 
in each of the steps of the attack. Determinism guarantees that the three state transitions occur in the attack. Firstly, the eviction of the target line(s) from the cache in  \emph{setup}, followed by 
insertion of a cache line in \emph{interact}. Finally, timing differences are observed based on the presence and/or absence of specific cache lines in \emph{receive}. Likewise, information flow and sharing guarantee that the secret is encoded and communicated, through \emph{MS} of the shared \emph{SC}, from the victim to the adversary, across the different steps. Access violation is not exploited in this attack since the \emph{interact} step, executed by the victim, does not lead to an access outside its own protection domain, \textcolor{black}{i.e., there is not an access violation on the microarchitectural level}. In general, attacks can exploit a subset or all the root causes as we will show in Section \ref{putting_it_all} and \ref{root_causes}.}

\vspace{-.2em}
\subsection{Threat Model}
\vspace{-.2em}
\label{sec:threat_model}
We consider four types of threat models in our classification. Across the different threat models, the secret, that the adversary attempts to steal, resides in a different protection domain from that of the adversary. 

An adversary can execute on a separate core from the victim, referred to as \emph{CrossCore}; be time-multiplexed on the same core as the victim process, referred to as \emph{SameThread};  run on distinct SMT threads executing on the same core, referred to as \emph{SMT} or run in isolation, referred to as \emph{Solo}. In the \emph{Solo} threat model the adversary only needs to have a pointer to the location of the victims data (kernel memory). The threat model determines which set of microarchitectural resources 
are shared or private in the 
attack setting on a given machine. A \emph{CrossCore} threat model leads to a scenario where fewer microarchitectural resources are 
shared.
In contrast, assuming the \emph{SameThread} or the \emph{SMT} threat model leads to potentially more microarchitectural resources being shared between 
victim and 
adversary, leading to a broader attack surface. 

Another dimension of the threat model is based on whether the adversary or the victim performs the different actions in an attack. In a typical attack the adversary performs one or more steps. However, it has been 
shown that an adversary can manipulate the victim to perform some of the required actions through the use of specific gadgets. This is especially useful in scenarios where the adversary does not have access to a shared microarchitectural resource state to facilitate the 
\emph{MS} transitions. This strategy 
increase the scope of possible attacks even in cases where the threat models limit the attack surface.

One example is the Spectre v2 attack~\cite{spectre} which requires training the Branch Target Buffer (\emph{BTB}) as part of the \emph{setup} step. Without 
gadgets 
such attacks would only be possible with the \emph{SMT}/\emph{SameThread} threat models since the \emph{BTB} is not shared between cores. However, when the victim can be manipulated to perform the training, a \emph{CrossCore} threat model can 
be used. The manipulation from the adversary can be performed by calling a function in the victims code with a controlled input, i.e., the action of triggering a gadget. The gadget can be constructed using Return Oriented Programming (ROP)~\cite{ROP07} where code snippets ending with a return instruction are used by changing the return address and thereby chaining the different snippets together. 
However, these attack scenarios depend on the availability of gadgets and/or vulnerabilities, such as buffer overflows, and most have not been demonstrated outside of specific environments~\cite{SysEvauationTransient19}. 

For some optimizations and attack scenarios the side-channel 
can be noisy and/or obscured by other optimizations, thereby making it difficult to decode the secrets based on \emph{MS} transition and the consequent timing variations. 
Amplification gadget(s) can be used by the adversary to enhance the timing differences and 
ease the decoding of the secret
. A simple example is on the cache side-channel, where prefetching can obscure the secret-related accesses. This can be circumvented 
by using a linked list~\cite{EffCacheAtt10} or by spreading accesses across pages~\cite{spectre} since most 
prefetchers only target linear or strided access patterns and do not prefetch across page boundaries. 


\vspace{-.2em}
\subsection{Case Study}
\vspace{-.2em}
\label{putting_it_all}
Next, we will describe an actual attack, Spectre v1, using the abstract model, the root causes we have identified and 
the threat model, as a framework.

\subsubsection{Spectre v1}  

\begin{figure}[b!]
\centering
\vspace{-1.em}
\includegraphics[width=0.44\textwidth]{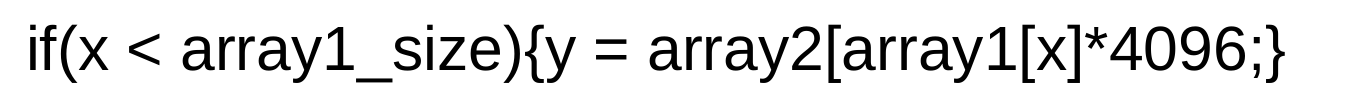}
\vspace{-1.em}
\caption{Example code for Spectre v1 attack~\cite{spectre}.}
\label{fig:spectre_pht_code_snippet}
\end{figure}

Spectre v1~\cite{spectre} leverages three different optimizations: branch prediction, speculative execution and shared cache. The example code for the attack is shown in Figure~\ref{fig:spectre_pht_code_snippet}, where \emph{x} is controlled by the adversary and is used to represent the address delta between the base address of \emph{array1} and the secret's location. The attack involves speculatively executing the if-clause code block, by training the branch predictor to predict taken. 
When the taken code block is speculatively executed the adversary can cause speculative access to \emph{array2} indexed using the adversary controlled 
\emph{x}. Even when the speculatively executed instructions are eventually rolled back this still leaves a trace in the cache, as a consequence of the access to \emph{array2}, which is then used to infer the secret.


\begin{figure}[t!]
\centering
\includegraphics[width=0.52\textwidth]{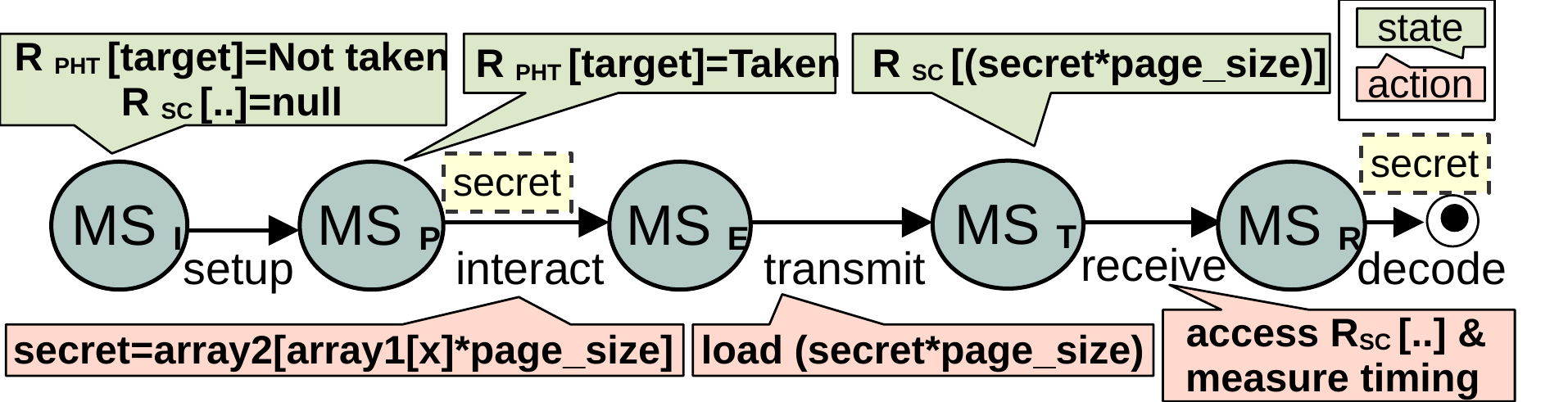}
\vspace{-0.5em}
\caption{\textcolor{black}{\emph{MS} transitions and actions in Spectre v1.}}
\label{fig:case_study}
\vspace{-1em}
\end{figure}

Figure~\ref{fig:case_study} show an overview of the attack steps and MS transitions. 
The \emph{setup} step consists of (mis)training the branch predictor Pattern History Table (\emph{PHT}) \textcolor{black}{by adversary/victim} to change the prediction for the targeted conditional branch from \emph{MS\textsubscript{I}\{R\textsubscript{PHT}[[target]=Not\ taken]\}} to \emph{MS\textsubscript{P}\{R\textsubscript{PHT}[[target]=Taken]\}}. 
The necessary (mis)training can either be performed by the adversary, restricting the threat model to SMT/SameThread, or be performed by the victim. To make the victim perform the (mis)training of the specific conditional branch a gadget must be located in the victim binary containing instructions which execute and train the target \emph{PHT} entry to mispredict on the conditional branch. Setup of the \emph{SC}, as in flush+reload, is also needed since it will be used later in the \emph{transmit} step.

In the \emph{interact} step the victim executes speculatively and accesses the secret because of the (mis)prediction. Speculative execution allows the CPU to temporarily violate program semantics by transiently execute code that accesses the secret and leave a trace in the \emph{MS}. In the \emph{transmit} step the secret is re-encoded in the state of another 
microarchitectural resource through a secret dependent access to the \emph{SC} (\emph{load\ (secret*page\_size)}). This access causes the \emph{SC} to transition to \emph{MS\textsubscript{T}\{R\textsubscript{SC}[(secret*page\_size)]\}}. The \emph{MS} transition occurs before the processor detects that the speculative execution was erroneous and rolls back the register state, leaving a trace in the state of the \emph{SC}. 
In the \emph{receive} step the adversary accesses the cache, measures the time and infers the secret based on 
timing variations for 
cache lines. 

The root causes which enable this attack are determinism, sharing, access violation 
and information flow. Determinism guarantees that, the adversary can cause a \emph{BTB} state update in an intended entry in the \emph{setup} step, which is then used in the \emph{interact} step. In addition, determinism ensures that the \emph{SC} becomes primed, as a result of the flush, in \emph{setup} and that the secret-dependent loads will result in timing variations corresponding to the lines presence/absence observed in the \emph{receive} step.
Access violation enables access to the secret which resides in a different protection domain, 
through speculative execution. Sharing is exploited in both the \emph{BTB} and the \emph{SC} during the \emph{setup} step and in the cache during the \emph{receive} step. Information flow is allowed through each of the state of the shared resources, across the different attack steps. 



\section{Systematization of Attacks}
\label{root_causes}
In this section we use the abstract model, the root causes and the threat model as a framework 
to systematize attacks on a broad set of microarchitectural optimizations. We 
describe a typical attack on each optimization and analyze the \textcolor{black}{necessary} root causes exploited in the different steps of the attack. We group attacks wherever possible and also discuss 
dissimilarities between 
the attacks on the same microarchitecture optimization. Note that the goal of this analysis is not to exhaustively discuss all the attacks proposed in literature. 
Rather, through the discussion
, our aim is to highlight commonalities and differences across different attacks that target a microarchitectural optimization by discussing the following questions: 
\begin{enumerate}
\vspace{-.2em}
    \item Which microarchitectural resource(s) are exploited in an attack?
    \item Which root cause(s) \textcolor{black}{are necessary to} enable the attack and in which attack step(s)?
    \item Under which threat model(s) is/are the attacks possible?
\vspace{-.2em}
\end{enumerate}

To simplify the discussion we categorize the microarchitectural optimizations into two broad groups --  non-transient and transient optimizations -- and analyze attacks on each of them.
The systematization of attacks is presented in Table \ref{table:new_attack_classification}. 
Finally, we discuss microarchitectural optimizations without any published attacks that are prone to leaks.

\begin{scriptsize}
\begin{table*}[h!]
\vspace{-.5em}
  \small
  \scalebox{0.8}{
  \renewcommand{\arraystretch}{1.16}
  \setlength\tabcolsep{1.5pt} 
  \centering
  \begin{tabular}{|l|c|c|c|c|c|c|c|c|c|}
    \hline
    \textbf{Microarchitectural}%
    &\textbf{Attack(s)} &\textbf{Resource(s)}  &\multicolumn{4}{c|}{\textbf{Attack steps}}&\textbf{Threat}\\
     \textbf{optimization} 
     & & & \textbf{$A_S$} & \textbf{$A_I$} & \textbf{$A_T$} & \textbf{$A_R$} &\textbf{model}  \\ 
    \hline
   \hline
  Shared cache (SC) & flush+flush~\cite{flush_flush}, flush+reload~\cite{flush_reload,CacheGames11,waitAMinute_14} & $R_{SC 
  }$ & D/S/I [A] & D/S/I [V] & - & D/S/I [A] & 1,2,3 \\ 
   & prime+probe~\cite{prime_probe,spy_in_sandbox_15,AS_15,cache_attack_16,evict_time_06}& $R_{SC
   }$ & D/S/I [A] & D/S/I [V] & - & D/S/I [A] & 1,2,3  \\
   & observation~\cite{Shusterman}, C5~\cite{C5_15} & $R_{SC
   }$ & D/S/I [A] & D/S/I [V] & - & D/S/I [A] & 1 \\
   & collide+probe~\cite{takeAWay_20} & $R_{SC
   }$ & D/S/I [A] & D/S/I [V] & - & D/S/I [A] & 2 \\
   & load+reload~\cite{takeAWay_20} & $R_{SC
   }$ & D/S/I [A] & D/S/I [V] & - & D/S/I [A] & 3 \\
  & xlate+probe/abort~\cite{Malicious_Management_Unit_18} & $R_{SC}$, $R_{MMU}$ & D/S/I [A] & D/S/I [V] & - & D/S/I [A] & 1,2,3 \\ 
   & \textcolor{black}{TLBleed~\cite{TLB_bleed_18,TLB_DR_22}} & $R_{SC}$, $R_{TLB}$ & D/S/I [A] & D/S/I [V] & - & D/S/I [A] & 2,3 \\
   &\textcolor{black}{CacheBleed~\cite{CacheBleedYarom2017}} & $R_{SC_{bank}^T}$ & D/S/I [A] & D/S/I [V] & - & D/S/I [A] & 3 \\ 
   & \textcolor{black}{MemJam~\cite{MEMJAM}} & $R_{SC}$ & D/S/I [A] & D/S/A/I [V] & - & D/S/I [A] & 2,3 \\ 
   & \textcolor{black}{LoR~\cite{LordoftheRings}} & $R_{ring^T}$ & D/S/I [A] & D/S/I [V] & - & D/S/I [A] & 1 \\ 
   & \textcolor{black}{MeshUp~\cite{MeshUp} MeshAround~\cite{MeshAround22}} & $R_{mesh^T}$ & D/S/I [A] & D/S/I [V] & - & D/S/I [A] & 1 \\ 
  & CacheTiming~\cite{Bernstein2005CachetimingAO,Miyauchi_2003} & $R_{SC}$ & - [A] & D/I [V*] & - & D/I [A] & 1/-\^2 \\
 \hline
 Prefetching (P) & Prefetch SCAs\cite{pref_side_channel_attacks_16,amdAttacksSpaceTime} & $R_{P}$,$R_{SC}$,$R_{TLB}$ & D/S/I [A] & D/S/A/I [V] & - & D/S/I [A] & 2,3 \\ 
 & prefetch+reload, prefetch+prefetch~\cite{pref_pref2022} & $R_{P}$,$R_{SC}$ & D/S/I [A] & D/S/I [V] & - & D/S/I [A] & 1 \\
 & LeakingControlFlow~\cite{chen2021leaking} & $R_{P}$,$R_{SC}$,$R_{TLB}$ & D/S/I [A] & D/S/A/I [V] & - & D/S/I [A] & 2,3 \\
 & DMP~\cite{augury22,Pandora} & $R_{P}$,$R_{SC}$ & D/S/I [V*/A] & D/S/A/I [V*/A] & - & D/S/I [A] & 2 \\
& Unveiling~\cite{unvelining} & $R_{P}$,$R_{SC}$ & D/S/I [A] & D/S/I [V] & - & D/I [A] & 1,2,3 \\
 & FetchingTale~\cite{fetchingTale} & $R_{P}$,$R_{SC}$ & D/S/I [A] & D/S/I [V] & - & D/S/I [A] & 2,3 \\
   \hline
  Branch  
  & JumpOverASLR~\cite{jumpOverASLR16} & $R_{BTB}$ & D/S/I [A] & D/S/A/I [V] & - & D/S/I [A] & 2,3 \\
  prediction (BP) & PredictingKeys~\cite{SimplePredictionAnalysis,predicitingkeys07,NewBP_vulnerabilities_07,jumpOverASLR16} & $R_{BTB}$ & D/S/I [A] & D/S/I [V] & - & D/S/I [A] & 2,3 \\
  & BranchScope~\cite{BranchScope18} & $R_{PHT}$ & D/S/I [A] & D/S/A/I [V] & - & D/S/I [A] & 2,3 \\
  & BranchShadowing~\cite{branch_shadowing_sgx_17} & $R_{PHT}$,$R_{LRB}$ & D/S/I [A] & D/S/A/I [V] & - & D/S/I [A] & 2,3 \\
  \hline
    Computational 
    & Subnormal FP~\cite{subnornalFP,Kohlbrenner17} & $R_{FPU}$ & (D/S/I)$^1$ [A] & D (S/I)$^1$ [V] & - & (D/S/I)$^1$ [A] & - \\ 
    simplification (CS) 
    & Early termination~\cite{early_termination_10} & $R_{MUL}$ & - [A] & D [V*] & - & - [A]  & 2,3 \\
\hline
      Transient attacks: 
      & Spectre v1~\cite{spectre,meltdownprime_18}, v1.1~\cite{kiriansky2018speculativeBUFFEROVERFLOWS} & $R_{SC}$,$R_{PHT}$,$R_{BHB}$ & D/S/I [A/V*] & D/S/I [V*] & D/S/A/I [V*] & D/S/I [A] & 1,2 \\
      Speculation-based 
      & Spectre v2~\cite{spectre,SgxPectre_19,BranchHistoryInjection22,RETBLEED_22} & $R_{SC}$,$R_{BTB}$,$R_{BHB}$ & D/S/I [A/V*] & D/S/I [V*] & D/S/A/I [V*] & D/S/I [A] & 1,2 \\
      & Spectre v4~\cite{spectre_is_here_to_stay_19}, LVI~\cite{LV1_20} & $R_{SC}$,$R_{STL}$ & D/S/I [A/V*] & D/S/I [V*] & D/S/A/I [V*] & D/S/I [A] & 2 \\
      & Spectre v5 (ret2spec)~\cite{spectre_returns_18,Ret2spec_18} & $R_{SC}$,$R_{RSB}$,$R_{BTB}$ & D/S/I [A/V*] & D/S/I [V*] & D/S/A/I [V*] & D/S/I [A] & 1,2 \\
    & BranchSpec~\cite{BranchSpec_20} & $R_{PHT}$ & D/S/I [A/V*] & D/S/I [V*] & D/S/A/I [V*] & D/S/I [A] & 2,3 \\
    & NetSpectre~\cite{netspectre_19} & $R_{PHT}$,$R_{AVX2}$/$R_{SC}$ & D/S/I [V*] & D/S/I [V*] & D/S/A/I [V*] & D/S/I [A] & -$^2$ \\
    & CROSSTALK~\cite{crosstalk_21} & $R_{SC}$,$R_{LFB}$,$R_{staging\_buf.}$ & D/S/I [A/V*] & D/S/I [V*] & D/S/A/I [V*] & D/S/I [A] & 1 \\
     & SMoTherSpectre~\cite{SMoTherSpectre_19} & $R_{SC}$,$R_{PHT}$,$R_{ports^T}$ & D/S/I [A] & D/S/I [V*] & D/S/A/I [V*] & D/S/I [A] & 3 \\
     & SpectreRewind~\cite{SpectreRewind_20} & $R_{SC}$,$R_{BTB}$,$R_{ports^T}$ & D/S/I [A] & D/S/I [V*] & D/S/A/I [V*] & D/S/I [A] & 2,3 \\
     & Speculative interference~\cite{Speculative_interference_attacks_21} & $R_{SC}$,$R_{BTB}$,$R_{MSHR}$,$R_{RS}$,$R_{EU^T}$ & D/S/I [A/V*] & D/S/I [V*] & D/S/A/I [V*] & D/S/I [A] & 1,2,3 \\
    & ROB cont.~\cite{Reorder_buffer_contention_21} & $R_{PHT}$,$R_{ROB}$ & D/S/I [A/V*] & D/S/I [V*] & D/S/A/I [V*] & D/S/I [A] & 1,2,3 \\
     \hline
     Transient attacks: 
     & Meltdown~\cite{Meltdown,meltdownprime_18} & $R_{SC}$,$R_{BTB}$ [PF-US] & D/I [A] & D/S/A/I [A] & D/I [A] & D/I [A] & 4 \\
     Exception-based 
     & Foreshadow~\cite{foreshadow_18,foreshadowNG_18} & $R_{SC}$,$R_{TLB}$ [PF-P] & D/I [A] & D/S/A/I [A] & D/I [A] & D/I [A] & 4 \\
     & Spectre1.2~\cite{kiriansky2018speculativeBUFFEROVERFLOWS} & $R_{SC}$,$R_{TLB}$ [PF-RW] & D/I [A] & D/S/A/I [A] & D/I [A] & D/I [A] & 4 \\
     & LazyFP~\cite{stecklina2018lazyfp} & $R_{SC}$,$R_{FPU}$,$R_{SIMD}$ [\#NM] & D/S/I [V] & D/S/A/I [A] & D/I [A] & D/I [A] & 2,3 \\
     & Fallout ~\cite{fallout_19} & $R_{SC}$,$R_{SB}$ & D/S/I [A] & D/S/A/I [A\&V] & D/I [A] & D/I [A] & 2,3 \\
     & RIDL~\cite{RIDL_19} ZombieLoad~\cite{ZombieLoad_mds_19} & $R_{SC}$,$R_{LFB}$ & D/S/I [A] & D/S/A/I [A\&V] & D/I [A] & D/I [A] & 2,3 \\
    & LVI~\cite{LV1_20} & $R_{SC}$,$R_{FPU}$,$R_{SB}$,$R_{LFB}$ [PF] & D/S/I [A] & D/S/I [V*] & D/S/A/I [V*] & D/S/I [A] & 2,3,4 \\
  \hline
  \end{tabular}}
  \caption{Attack systematization. 
  Root causes: determinism (D), sharing (S), access violation (A), information flow (I). Attack steps: setup ($A_S$), interact ($A_I$), transmit ($A_T$), receive ($A_R$), performed by adversary [A] or victim [V]. Gadget*. Threat models: 1) CrossCore, 2) SameThread, 3) SMT, 4) Solo.$^1$In SW.$^2$Remote.$^T$Temporal resource.}
  \label{table:new_attack_classification}
  \vspace{-2.em}
\end{table*}
\end{scriptsize}

\subsection{Non-transient attacks}

\subsubsection{Cache}
\label{attack_cache}
The last-level cache is typically shared among cores to improve cache utilization and to reduce costly off-chip accesses. The \emph{SC} 
is vulnerable to side-channel attacks and is an attractive attack surface because of the channel characteristics (i.e., low noise and high attack bandwidth~\cite{Sibert_theintel_95}). There exist three high-level attack categories: i) reuse-based~\cite{flush_flush,flush_reload,CacheGames11,waitAMinute_14} where data is shared between adversary and victim allowing both to access it, ii) 
conflict-based~\cite{prime_probe,spy_in_sandbox_15,AS_15,cache_attack_16,evict_time_06,takeAWay_20}, where an adversary creates conflicts to evict \textcolor{black}{target} lines belonging to the victim, and iii) observation-based~\cite{Shusterman,cache_template_15}, \textcolor{black}{i.e., brute-force conflicts.} 
\textcolor{black}{In observation-based attacks the conflicts and observed behaviour relates to any cache line and is not restricted to a selected target, as in conflict-based attacks}. 
\textcolor{black}{We note that the same categories are common across attacks on other shared resources and have implications for the defense strategies (see Section \ref{systematic_defenses}).} 



Prime+probe~\cite{prime_probe}, a typical conflict-based attack ~\cite{prime_probe,spy_in_sandbox_15,AS_15,cache_attack_16,evict_time_06,takeAWay_20}, is used in the absence of data sharing between an adversary and a victim.   
The threat model is \emph{CrossCore}, since the \emph{SC} is shared between cores. In the \emph{setup} step, the adversary first finds the set of cache lines (eviction-set) which will create conflicts in the targeted index shared with cache lines belonging to the victim and evict them from the cache. The state changes to \emph{MS\textsubscript{P}\{R\textsubscript{SC}[[index]=A\textsubscript{adversary}]\}}. Next, in the \emph{interact} step, the victim accesses the secret during execution which is in turn encoded in the \emph{SC} state through the presence/absence of the specific target cache line(s). The state changes to \emph{MS\textsubscript{E}\{R\textsubscript{SC}[[index]=B\textsubscript{victim}]\}}. In the \emph{receive} step the adversary accesses the target cache line(s) and based on the timing variations infers the secret. 

The root causes determinism, sharing and information flow enable the attack, as shown in Table \ref{table:new_attack_classification}. Determinism provides different guarantees in the three steps of the attack. First, in the \emph{setup} step it ensures that conflicts can be created at a specific index in the cache which causes an eviction of the targeted cache line. Second, in the \emph{interact} step it ensures the secret is encoded through the \emph{MS} of the \emph{SC}. Third, it also causes timing variation corresponding to the presence/absence of the cache line. Sharing and information flow ensures that the secret is encoded and communicated, through \emph{MS} of the \emph{SC}, from the victim to the adversary, across the different steps.





\textcolor{black}{Observation-based attacks~\cite{Shusterman,cache_template_15} work by observing set conflicts in all the sets in the cache instead of actively causing them in a few sets like prime+probe. These attacks leverage the observation that the same conflict patterns will re-occur because the cache behavior of a program is deterministic. An alternative \emph{setup} step is used, without relying on using \emph{clflush} or a specific eviction-set, which involves accessing a large buffer~\cite{Shusterman,C5_15,netspectre_19} to flush all cache lines in the \emph{SC} and creating conflicts across all the sets in the cache. 
This approach, however, has the drawback of a lower bandwidth since filling the \emph{SC} is time consuming. 
These observation-based attacks are challenging to defend against since they rely on determinism and the shared \emph{MS} of the \emph{SC} in the \emph{interact} and \emph{receive} steps. This limits the defenses which can be used to avoid the attack (for details see Section \ref{def_cache}).} 

A number of attacks show that other state 
can also be used for attacks, such as the state of replacement metadata~\cite{DAWG_MICRO}, way prediction~\cite{takeAWay_20}\textcolor{black}{, interconnect~\cite{MeshUp,LordoftheRings}, cache banks~\cite{CacheBleedYarom2017}, translation lookaside buffer (\emph{TLB})~\cite{TLB_bleed_18,TLB_DR_22}} or memory management unit (\emph{MMU})~\cite{Malicious_Management_Unit_18}. Using other microarchitecture resource state (related to \emph{SC}) 
enables to circumvent defenses on the \emph{SC}.   
Van Schaik et al.~\cite{Malicious_Management_Unit_18} 
propose a prime+probe-like attack 
where an eviction-set is found and used in the \emph{MMU} instead of the \emph{SC} to overcome defenses targeting conflict-based attacks in the \emph{SC}. \textcolor{black}{Wan et al.~\cite{MeshUp} show an attack using temporal contention on the mesh interconnect, while Paccagnella et al.~\cite{LordoftheRings} show an attack on the ring interconnect.} These attacks exploit determinism, sharing and information flow as previously discussed in the \emph{SC} attacks. \textcolor{black}{However, the sharing is of other microarchitectural resources as the \emph{MMU} or the mesh interconnect, instead of the \emph{SC}. We do not exhaustively cover all \emph{SC} related attacks since our goal is to discuss a representative set to demonstrate the applicability of our framework.} 







\subsubsection{Prefetching}
Prefetchers predict addresses that will be used by a program and proactively fetches them to help hide memory access latency. The attacks exploiting prefetching can be broadly grouped into two \textcolor{black}{fundamentally different} categories, 1) SW-based and 2) HW-based. Attacks that belong to the latter category exploit the availability of a HW prefetcher (microarchitectural resource) and specifically utilize the \emph{MS} of the prefetch tables and/or the \emph{SC} as the side channel while attacks that belong to the former category exploits different prefetch instructions directly \textcolor{black}{which exhibits different timing depending on the} state of the \emph{TLB} and/or the \emph{SC}
. The attacks exploiting SW-based and HW-based prefetching mechanisms can be further subdivided into two categories, 1A and 2A) Attacks which exploit the lack of permissions check and 1B and 2B) Attacks which exploit a secret-dependent prefetching pattern.

One typical attack from category 1A, a SW-based attack that exploits the lack of permissions check, is the address-translation attack by Gruss et al.~\cite{pref_side_channel_attacks_16} where the goal of the full attack is to translate between virtual and physical addresses from unprivileged user-space and overcome the protection provided by user-space and kernel-space Address Space Layout Randomization (ASLR). 
We focus on the first phase of the attack where the adversary searches through possible addresses and tests if two virtual
addresses, $a$ and $a'$, map to the same physical address by performing the attack. Here, address $a'$ can be a kernel address or a non-mapped address and not be directly accessible to the adversary. 
In the \emph{setup} step the adversary flushes the candidate collision address, $a$. The state changes to \emph{MS\textsubscript{P}\{R\textsubscript{SC}[[index\textsubscript{a}]=empty]\}}. In the \emph{interact} step, the adversary prefetches the address $a'$, and performs an access violation. The access violation 
is due to speculative dereferencing of kernel-space registers from user-space~\cite{Derefencing21}, and not because of the prefetch instruction as 
suggested by Gruss et al.~\cite{pref_side_channel_attacks_16}. 
The state changes to 
\emph{MS\textsubscript{E}\{R\textsubscript{SC}[[index\textsubscript{a}]=a']\}}. In the \emph{receive} step the adversary accesses 
address $a$ and based on the timing variations infers if there is a match between $a$ and $a'$. 




The root causes exploited by the attack are determinism, sharing, access violation and information flow, as shown in Table \ref{table:new_attack_classification}. Determinism enables all the three steps of the attack. First, in \emph{setup} step, it causes the cache line corresponding to $a$ to be evicted. Second, in the \emph{interact} step, it causes the prefetch of $a'$ to be encoded through the \emph{MS} of the \emph{SC}. Third, it makes timing variation to be observed corresponding to the presence/absence of the cache line. Access violation enables the attack by permitting the adversary to prefetch inaccessible address(es). Sharing and information flow guarantees that the secret is encoded and communicated, through \emph{MS} of the \emph{SC}.

In the second category in SW-based attacks, 1B, the attacks~\cite{pref_pref2022} use a SW-controlled prefetch instruction (\emph{PREFETCHW}), 
in the \emph{setup} and the \emph{receive}, to reveal cryptographic keys through the data-dependent access pattern of an application. 




The attacks in category 2A, HW-based without permissions check, 
use HW prefetchers to prefetch addresses outside of a sand-box~\cite{augury22} or in kernel-space~\cite{chen2021leaking}. The prefetcher is trained in the \emph{setup} step, in order to issue a prefetch to the target address in the \emph{interact} step. Chen et al.~\cite{chen2021leaking} show that an 
adversary trained prefetcher can prefetch kernel addresses. 
In ~\cite{augury22} a data memory-dependent prefetcher (DMP) is trained to perform out-of-bounds reads on pointers, since the prefetcher is allowed to use memory content to prefetch irregular address patterns. The root causes exploited by the attacks are determinism, sharing, access violation and information flow. In both the attacks the state of the \emph{SC} is used to encode the accesses in the \emph{interact} step and to measure the timing difference in the \emph{receive} step. \textcolor{black}{There is no need of an additional \emph{transmit} step to encode the secret in the \emph{SC}, since the prefetcher directly interacts with the \emph{SC} in the \emph{interact} step.} Access violation is exploited in the \emph{interact} step since the prefetch is issued without 
permission checks.

Lastly, the attacks~\cite{unvelining,chen2021leaking} in category 2B, leak secrets through secret data-dependent prefetching pattern(s). Shin et al.~\cite{unvelining} show how data-dependent prefetch activity can be used to leak secret keys through \emph{MS} in the \emph{SC}. 
Similar to the previously discussed attacks using SW-based prefetching, which also leak secrets through data-dependent prefetch access patterns, the root causes are determinism, sharing and information flow.

\subsubsection{Branch prediction}
Branch predictors record history of branch outcomes in order to predict the direction of control flow after a branch instruction, 
to improve 
instruction flow. There are two high-level strategies for attacks using the branch predictor, reuse-based~\cite{BranchScope18,branch_shadowing_sgx_17,spectre} 
where entries set by one process may influence the other and conflict-based~\cite{jumpOverASLR16,SimplePredictionAnalysis,NewBP_vulnerabilities_07,predicitingkeys07} where contention is used to evict the entry inserted by the other process.  





A typical conflict-based attack is JumpOverASLR~\cite{jumpOverASLR16} where the goal of the adversary is to determine the position of a code block in the address space of a victim. Knowing the position of a code block can help break the protection provided by ASLR since the randomization is based on an offset. The attack is launched multiple times using different index values, searching for a collision in the 
\emph{BTB}. A collision in the \emph{BTB} can be used to infer the address used by the victim and the offset used by ASLR. In the \emph{setup} step, the adversary inserts an entry in the \emph{BTB} which might create a collision with the entry later inserted by the victim code. This changes the \emph{MS} to primed \emph{MS\textsubscript{P}\{R\textsubscript{BTB}[[index\textsubscript{I}]=addr.\ A]\}}. Next, in the \emph{interact} step the victim executes and inserts a different target address at the same \emph{BTB} position, creating a collision. The state transitions to \emph{MS\textsubscript{E}\{R\textsubscript{BTB}[[index\textsubscript{I}]=addr.\ B]\}}. In the \emph{receive} step the adversary executes code which will trigger the \emph{BTB} entry at \emph{index\textsubscript{I}} and measures the execution time. If the \emph{BTB} entry was changed by the victim  it would result in a longer execution time since the target address is incorrect ($B$ instead of $A$). 

The root causes which enable this attack 
are determinism, sharing and information flow, as shown in Table \ref{table:new_attack_classification}. Determinism guarantees that, the adversary can cause a \emph{BTB} state update in an intended entry, induce a conflict on the same entry when the victim executes and measure timing variations 
due to the conflict. Likewise, information flow and sharing guarantees that the secret is encoded and communicated, through \emph{MS} of the \emph{BTB}, from the victim to the adversary, across the different steps. Note that this attack is restricted to the \emph{SameThread} threat model (although it can be applied in \emph{SMT}) and does not extend to \emph{CrossCore} because \emph{BTB} state is not shared across cores. 

There also exist conflict-based attacks which exploit that the branch predictions can reveal data-dependent control flow~\cite{SimplePredictionAnalysis,NewBP_vulnerabilities_07,predicitingkeys07}. For example in ~\cite{predicitingkeys07} secrets are inferred based on the predictions made in the \emph{BTB}. 

The reuse-based attack use the 
\emph{PHT} instead of the \emph{BTB}~\cite{BranchScope18}. In BranchScope~\cite{BranchScope18}, the branch predictor is manipulated into using only the directional branch predictor, \emph{PHT}, where the directional prediction 
inserted by the adversary is changed by the victim which reveal the direction of conditional branches. The attack can also be used against SGX enclaves, since the \emph{PHT} is shared between processes executing in SGX and outside. The same root causes as in JumpOverASLR enable the attack.  

\subsubsection{Computational simplification}
Computational simplification comprises techniques which eliminate or simplify instruction execution. One example is the zero-skip multiplier and the same principle can be applied on different instruction types as square root, AND/OR and to different pipeline stages. Attacks on this type of optimizations have been studied~\cite{practicalMitig, subnornalFP, early_termination_10}. In ~\cite{subnornalFP} an attack is described using subnormals in a floating-point division unit, to create visible timing differences. 
Großschädl et al. ~\cite{early_termination_10} describe an attack using early-termination of multiplication where the multiplication is terminated when all remaining digits are zero, creating observable timing differences. The goal of the attack is to leak secret keys from cryptographic SW such as RSA. There are two prerequisites of the attack, firstly, that the adversary is able to control the plaintext which will be encrypted and secondly, that the timing can be observed on a side-channel. In the \emph{setup} the adversary calls the cryptographic function on the victim with a plaintext. In the \emph{interact} step the victim encrypts the plaintext and will experience different timings depending on the values of the key. Großschädl et al. does not describe which side-channel could be used in order to allow the adversary to observe the timing difference. We note that either the \emph{MS} of the \emph{SC} or execution unit contention could be used. A gadget is likely needed at the victim for re-encoding of the secret to the side-channel.

The root cause exploited 
is determinism which enables the data-dependent behaviour of the early-termination optimization and the timing variability. In addition, the side-channel, which enables the adversary to observe the timing differences, exploits sharing and information flow.





\vspace{-.2em}
\subsection{Transient attacks}
\vspace{-.2em}
\label{transient}
We finally discuss transient execution attacks which 
exploit speculative out-of-order (OoO) execution to execute code transiently (i.e. executed but never committed). We use the categorization provided in related works ~\cite{SysEvauationTransient19}, which divide the transient attacks broadly into two groups, speculation-based~\cite{spectre,SgxPectre_19,spectre_is_here_to_stay_19,Ret2spec_18,BranchSpec_20,netspectre_19,crosstalk_21,SMoTherSpectre_19,SpectreRewind_20,Speculative_interference_attacks_21,Reorder_buffer_contention_21} and exception-based~\cite{Meltdown,meltdownprime_18,foreshadow_18,kiriansky2018speculativeBUFFEROVERFLOWS,stecklina2018lazyfp,LV1_20,fallout_19,RIDL_19} attacks. 

\subsubsection{Speculation-based Attacks}
\label{speculation_based}
The attacks that fall in this category exploit transient execution, due to branch prediction and/or address/value speculation, to access the secret. An overview of the attacks is shown in Table \ref{table:new_attack_classification}. Spectre v2~\cite{spectre} is a typical speculation-based attack. 
The attack exploits an indirect branch to execute  a gadget which interacts with the secret in the victims protection domain, leaving a trace in the \emph{MS}. The prerequisites 
are an indirect branch that can be (mis)trained and a known gadget in the victim's binary that can be manipulated to interact with the secret. The 
threat models are \emph{SameThread} and \emph{SMT}, since \emph{BTB} is a resource private to a core. However, \emph{CrossCore} can be used if a gadget is used to make the victim perform the (mis)training. 
In the \emph{setup} step the adversary\textcolor{black}{/victim} (mis)trains the \emph{BTB} to insert a new entry containing the address of the gadget for the indirect branch. The state changes to \emph{MS\textsubscript{P}\{R\textsubscript{BTB}[[index\textsubscript{target}]=addr.\ A\textsubscript{gadget}]\}}. The root causes 
in the \emph{setup} step are determinism, sharing and information flow. Determinism guarantees that the adversary can cause a \emph{BTB} state update in an intended entry, while sharing 
and information flow enables the state change caused in the \emph{BTB} 
to be observed by the victim. Note that setup of the \emph{SC} is also performed, i.e. \emph{clflush}, since it will be used later in the \emph{transmit} step.

Next, in the \emph{interact} step the victim executes the gadget speculatively, accesses the secret and changes the \emph{MS}. The root causes are determinism, access violation and information flow. Determinism guarantees that the (mis)trained \emph{BTB} entry is used. Access violation enables access to the secret through execution of the gadget which results in temporary violation of program semantics, i.e., instructions that access the secret are executed and are later squashed. In the \emph{transmit} step the secret is re-encoded in the state of the \emph{SC}, by issuing load(s) to the target address(es) \textcolor{black}{by the victim}. The root causes exploited are determinism, sharing and information flow since the \emph{SC} contains the cache line(s) and the \emph{MS} of \emph{SC} is shared between the adversary and the victim. Finally, in the \emph{receive} step the adversary accesses the target cache line(s) and based on the timing variations infers the secret. The root causes are the same as in the previous step, with the difference that determinism 
ensures observable timing variations 
based on the state of the \emph{SC}, i.e., the presence/absence of the target cache line(s).


In contrast to Spectre v2, which uses the \emph{BTB}, other microarchitectural resources have been used to manipulate the control flow, e.g., \emph{PHT}~\cite{spectre} or the Return Stack Buffer (\emph{RSB})~\cite{spectre_returns_18,Ret2spec_18}. In addition, address speculation can be targeted for manipulating Store-To-Load (\emph{STL}) forwarding that happens in the Load Store Queue (\emph{LSQ})~\cite{spectre_is_here_to_stay_19}. Many of these different attack variants still use \emph{SC} as the side-channel for transmission. 

Next, we discuss attacks which are more restrictive since other microarchitectural resource(s) (not \emph{SC}) is/are used for the transmission of the secret. One example is BranchSpec~\cite{BranchSpec_20} where the \emph{PHT} is used in the \emph{transmit} and \emph{receive} step. The root causes exploited by the attack are the same as in Spectre v2, while the threat model is more restrictive since the \emph{PHT}, used as the side-channel, is not shared between cores. Another attack, SMoTherSpectre \cite{SMoTherSpectre_19},  uses port contention to encode the secret and transmit to a co-running SMT thread. Likewise, temporal contention in the floating-point division unit is exploited in SpectreRewind~\cite{SpectreRewind_20}. 
Like SpectreRewind, the attack proposed by Behnia et al.~\cite{Speculative_interference_attacks_21} shows that the secret can be encoded by affecting the timing and order of older instructions, which are issued before the secret dependent instruction(s) in program order. This is in contrast to prior works \cite{SafeSpec_19,InvisiSpec_18,CondSpec_19} that focused on studying the secret-dependent effect on younger instructions, issued after the secret dependent instruction(s). 
In the attack, the non-speculative instructions timing is affected either through the miss status handling register (\emph{MSHR}) or execution unit contention. As an example, let's consider the attack using the \emph{MSHR} described by Behnia et al.~\cite{Speculative_interference_attacks_21}. In the \emph{setup} step the adversary evicts a number of cache lines $Y$. In the \emph{interact} step the a gadget, depending on the secret value, either issues independent loads to the cache lines $Y$ (filling up the \emph{MSHR} entries) or issues loads to the same cache line (using one entry in the \emph{MSHR}). When the target victim load occurs it will experience different timing depending on the \emph{MS} of the \emph{MSHR}.

\subsubsection{Exception-based Attacks}
\label{exception_based}
The attacks that fall in this category exploit transient execution, due to delayed exception handling 
 to access the secret. 
Meltdown~\cite{Meltdown} is a typical attack from this category where the adversary exploits transient execution due to delayed exception handling, to read arbitrary kernel memory. In the \emph{setup} step the adversary causes an exception by accessing a kernel address that resides in a kernel memory page without suitable permissions causing a page fault, e.g., PF-US. Because of the deferred exception handling the execution continues transiently. In the \emph{interact} step the adversary executes code that uses the loaded value from the faulting kernel address. \textcolor{black}{By suppressing the exception can the transient execution continue
~\cite{Meltdown}}. In the \emph{transmit} step the secret is encoded in the \emph{MS} of the \emph{SC}, through a load to the data buffer, in order for the adversary to retain the information after the transient execution is rolled back after exception handling. In the \emph{receive} step the secret is inferred from the state of the \emph{SC}, through the presence/absence of cache line(s). Note that all the steps of the attack are performed by the adversary. 

The root causes enabling the attack 
are determinism, sharing, access violation and information flow (Table \ref{table:new_attack_classification}). Determinism ensures transient execution due to delayed exception handling in the \emph{setup} step and that the secret is encoded in the state of the \emph{SC} in the \emph{interact} step. Sharing enables the access from the adversary to the victim kernel address in the \emph{interact} step. Information flow enables the transiently accessed secrets to be communicated to the non-transient execution, through the \emph{MS} of the \emph{SC}. Access violation enables the adversary, in the \emph{interact} step, to access kernel data which it does not have the right privileges to access. 

Attacks have shown that different types of exceptions, i.e., page fault (PF), can be used in the \emph{setup} step. For instance, Foreshadow uses PF-P~\cite{foreshadow_18,foreshadowNG_18},  Spectre v1.2~\cite{kiriansky2018speculativeBUFFEROVERFLOWS} uses PF-RW while LazyFP~\cite{stecklina2018lazyfp} uses \#NM (device not available). Other types of page fault exceptions can also be used as shown by Canella et al.~\cite{SysEvauationTransient19}. Furthermore, in addition to reading from kernel memory, attacks have also exploited delayed exception handling to leak data across addresses spaces, virtual machines and even from secure enclaves ~\cite{foreshadow_18,foreshadowNG_18}.


Another group of attacks, referred to as the microarchitectural data sampling (MDS) attacks, exploit the state of internal buffers in the CPU, as the Line Fill Buffers (\emph{LFBs})~\cite{RIDL_19,ZombieLoad_mds_19} or the Store Buffer (\emph{SB})~\cite{fallout_19}, in conjunction with delayed exception handling. Specifically, the attacks leverage the observation that values from these buffers can be leaked as a consequence of accesses that trigger an exception. In the ZombieLoad v1 attack~\cite{ZombieLoad_mds_19} 
the adversary uses the kernel virtual address ($k$) corresponding to the user-space  address of the victim ($u$) where the secret resides. Both virtual addresses $k$ and $u$ map to the same physical address $s$. In the \emph{setup} step, the adversary monitors the victim by performing repeated flush+reload attacks on the address corresponding to the instruction just before the loading of the secret. This enables the attacker to synchronize with the victim and determine when it can start accessing the state of the buffers to retrieve the secret. In addition, the contents of the data buffer are also flushed from the \emph{SC}. Next, in the \emph{interact} step, the victim performs the load of the secret key, \emph{load\ u}. This load operation will 
cause the secret to be inserted in the \emph{LFB} (on a cache miss). The adversary performs a faulting load i.e. the ZombieLoad, to the kernel address of the secret (\emph{load\ k}), which 
causes the adversary to retrieve the secret from the \emph{LFB}. In the \emph{transmit} step the adversary uses the secret as an index to a data buffer to encode the secret into the \emph{MS} of the \emph{SC}, before the transient execution is rolled back. In the \emph{receive} step the adversary accesses the data buffer entries to infer the secret based on the timing variations arising due to \emph{SC} state. In contrast to Meltdown, victim's accesses cause the secret to be inserted in the internal buffers which are then leaked by loads that trigger exceptions.
All four root causes enable this attack.
Determinism enables all the steps of the attack while sharing the \emph{MS} of the \emph{LFB} allows for information flow between the victim and the adversary. Access violation occurs during  transient execution when the adversary is allowed to read stale data from the \emph{LFB}. Unlike the MDS attacks discussed previously, Load Value Injection (LVI)~\cite{LV1_20} uses the different types of exception-based vulnerabilities to inject data/code and control victim's execution by controlling the values in the internal CPU buffers. This attack exploits the same root causes as the MDS attacks discussed previously. 

\vspace{-.2em}
\subsection{Vulnerable optimizations}
\vspace{-.2em}
Several microarchitectural optimizations available in literature have not been reviewed in detail with an emphasis on security. 
We discuss possible attacks using our framework for one
such representative optimization, value prediction. We also analysed vulnerabilities in a few other optimizations, including the ones explored by Vicarte et al.~\cite{Pandora}, 
but omit them 
due to page constraints.
\label{attack_value_prediction}
\textit{Value prediction:}
This is a speculative optimization that aims to increase instruction-level parallelism (ILP) and hide memory access latency by predicting values for load misses and consequently breaking instruction dependencies~\cite{VP_96,load_VP_96,EOLE_VP_16,LoadVP_via_address_17,Seznec2018ExploringVP}. Accurate predictions can improve ILP by increasing the overlap between memory access(es) and useful computation(s) while mispredictions lead to pipeline squashes and re-execution of instruction(s). In a nutshell, value prediction is implemented using table-based structures and samples history to enable prediction.  

Both reuse-based and conflict-based attacks are possible, similar to the attacks described for the branch predictor. One possible attack that exploits the reuse behavior is to let the victim train the predictor leading to the secret being encoded in the predictor state.
The adversary can then trigger a prediction and use this to infer the secret.
Another possible attack strategy is to let the adversary (mis)train the predictor in order to induce the victim to access a secret (cause an access violation) using a gadget, akin to injection attacks. This can then be leaked to the adversary through a side-channel, such as the \emph{SC}.
Finally, conflict-based attacks could also be mounted by using secret dependent predictor use behavior and monitoring the state of the prediction tables to infer secrets.
The root causes exploited by the aforementioned potential attacks
are determinism, sharing and information flow. Determinism permits the \emph{MS} of the prediction table to be accessed and manipulated depending on the attack requirements. Sharing permits the \emph{MS} of the prediction tables to be accessible to both the adversary and the victim and enables information flow between adversary and victim. Access violation could also be exploited if the predictions cause the adversary and/or the victim to access data from outside the intended protection domain.

\textcolor{black}{In summary, we observe that the necessary conditions for most attacks are determinism, sharing and information flow and that there are few variations in the combinations of root causes. We also note that the goal of the attacks is either to i) exploit data-dependent implementations to  leak encryption keys, or ii) circumvent privilege checks to typically read kernel data. All attacks in the later category exploit access violation.}

\vspace{-.3em}
\section{Systematization of Defenses}
\vspace{-.3em}
\label{systematic_defenses}


We present a systematization of defenses against attacks targeting different microarchitectural optimizations. We specifically discuss optimizations for which several attacks and defenses exist in literature: cache, prefetching, branch prediction, computational simplification and transient execution attacks. For each of the defenses, we discuss which root cause(s) and the attack step(s) the defenses target. \textcolor{black}{Eliminating any of the root causes, exploited in a specific attack, in any of the attack steps can provide protection. Table \ref{table:new_attack_classification} show the different root causes for each attack, which can be targeted by a defence. } 
We categorize defenses into groups, wherever possible, in case there are similarities. In addition, we also describe the protection level offered against the discussed attacks using the optimization and the threat model targeted by the defense. Our goal is not to exhaustively cover defenses against all possible attacks targeting a microarchitectural optimization. Rather, it is to explore broad defense strategies,  
in which attack step and root cause they can be applied and their limitations. 

\vspace{-.3em}
\subsection{Defenses against Non-transient attacks}
\vspace{-.3em}
\subsubsection{Cache}
\label{def_cache}

\begin{scriptsize}
\begin{table}[b!]
 \vspace{-1.5em}
  \small
  \scalebox{0.81}{
  \renewcommand{\arraystretch}{1.16}
   \setlength\tabcolsep{1.5pt} 
  \centering
  \begin{tabular}{|l|c|c|c|c|c|c|c|c|c|}
    \hline
    \textbf{Defense} &\textbf{Res.}  &\multicolumn{4}{c|}{\textbf{Attack step}}&\textbf{Threat}&\textbf{P}\\
     &\textbf{} & \textbf{$A_S$} & \textbf{$A_I$} & \textbf{$A_T$} & \textbf{$A_R$}&\textbf{model} & \\ 
    \hline
   \hline
   disable clflush~\cite{flush_reload,SHARP}  & $R_{SC}$ & D/S/I & - & - & - & 1,2,3 & $\LEFTcircle$ \\
   part.:~\cite{HYBCACHE_20,IVcache_21,RP_PLcache,CATalyst} & $R_{SC}$ & D/S/I & D/S/I & - & D/S/I & 1,2 & $\LEFTcircle$ \\
   \hspace{7mm}~\cite{STEALTHMEM,secDCP,FairSDP,jumanji} & $R_{SC}$ & D/S/I & D/S/I & - & D/S/I & 1,2 & $\LEFTcircle$ \\
   part.: static~\cite{DAWG_MICRO,MI6,IRONHIDE} & $R_{SC}$ & D/S/I & D/S/I & - & D/S/I & 1,2 & $\CIRCLE$ \\
   rand.~\cite{newcache,ceasar18,DAS18,CEASAR-S} & $R_{SC}$  & D/I & D/I & - & D/I & 1,2 & $\LEFTcircle$\\
   \hspace{7mm}~\cite{Tan2020PhantomCacheOC,scattercache} & $R_{SC}$  & D/I & D/I & - & D/I & 1,2 & $\LEFTcircle$\\
   repl.~\cite{RIC_17,Mirage,RandomFillCache,RP_PLcache,nonDeterministicCaches,NoMo} & $R_{SC}$ & D/I & D/I & - & D/I & 1,2 & $\LEFTcircle$ \\
   const. time~\cite{constant_time05,constant_time06,practicalMitig} & - & - & D/I & - & - & 1,2,3 & $\CIRCLE$\\
 \hline
  \end{tabular}
  }
  \caption{Defenses for attacks using the \emph{SC}. Protection (P): full$\CIRCLE$/partial$\LEFTcircle$.}
  \label{table:cache}
  \vspace{-1.5em}
\end{table}
\end{scriptsize}

Several defenses have been proposed for securing the shared cache against side-channel attacks. We have classified the different defenses for the \emph{SC} based on the root cause(s) and attack step(s) they target, into five broad categories, see Table \ref{table:cache}. Each row in the table shows which root causes are restricted by the defence, in which attacks step using which microarchitectural resource.

Disabling \emph{clflush}~\cite{flush_reload,SHARP} only addresses reuse-based attacks which typically uses the instruction in the \emph{setup} step. This affects 
the three root causes determinism, sharing and information flow primarily in the \emph{setup} step. 

The defenses in the next category, partitioning (part.), 
target 
sharing and information flow, in all the attack steps, by providing isolation between processes/threads in the \emph{SC} state using partitioning ~\cite{RP_PLcache,CATalyst,STEALTHMEM,secDCP,FairSDP,jumanji,DAWG_MICRO,MI6,IRONHIDE,optimus,BespokeCE,IVcache_21}. Partitioning can, in principle, provide full protection against reuse-based, conflict-based and observation-based attacks \textcolor{black}{where victim and adversary share the \emph{SC} state.} 
However, the defenses provide different levels of protection depending on the level of isolation, i.e., whether 
all or only some of the data is partitioned. For example, MI6 and IRONHIDE~\cite{IRONHIDE,MI6} statically partition both \emph{SC} and DRAM and can thereby enable full protection of the \emph{SC}, albeit at a comparatively higher performance cost. In contrast, STEALTHMEM~\cite{STEALTHMEM} only provides partial protection for a limited number of cache lines per core, which are not allowed to be evicted. This will lead to higher performance but also a lower protection level, since the state of the unprotected cache lines are shared. 



The next category, 
randomization (rand.), targets conflict-based attacks and is typically achieved by modifying the mapping of addresses to sets 
in the \emph{SC}~\cite{newcache,ceasar18,DAS18,CEASAR-S,Tan2020PhantomCacheOC,scattercache}.
Randomization target determinism and information flow. The 
strategy 
affects determinism by complicating the process of creating an eviction-set needed to evict a target cache line at a specific address. 
The change in mapping leads to limited information flow. This makes randomization effective especially against conflict-based attacks. 
However, in spite of defense, the \emph{SC} is still shared and insertions by the adversary can result in evictions for the victim process and vice versa. This limits the effectiveness against observation-based attacks since the working-set size of an application can be observed and can be leveraged by attacks~\cite{cacheFX}. 

The category replacement-based defenses (repl.) -- insertion and/or eviction -- leverage randomization to provide protection ~\cite{RIC_17,Mirage,RandomFillCache,RP_PLcache,nonDeterministicCaches,NoMo}. These proposals mainly target determinism and information flow through the \emph{SC} state. Specifically, through randomising insertion and/or eviction, 
the \emph{SC} 
do not react in the same way under the same preconditions.
Information flow is limited by reducing/avoiding set conflicts, i.e., by preventing an adversary from evicting data inserted by the victim. These defenses offer protection but eventually leak information 
since determinism, sharing and information flow in the \emph{SC} are not completely eliminated~\cite{Szefer2019}. 


Lastly, constant-time programming paradigm~\cite{constant_time05,constant_time06,practicalMitig} can be used 
to avoid data-dependent implementations which affects determinism and information flow. 
However, this is challenging to utilize in practice since it cannot be generically applied.




\subsubsection{Prefetching}

Existing defenses to protect against prefetch-based attacks can be categorized broadly into five groups, as shown in Table \ref{table:prefeching}. 
Disabling the prefetcher~\cite{unvelining,amdAttacksSpaceTime,fetchingTale,chen2021leaking} impacts 
determinism, sharing, access violation and information flow, in all the attack steps
. This strategy is equally applicable to attacks using 
HW- or SW-based prefetching. 
However, the performance cost can be high since prefetching can provide 
significant speedups. 


The defense in the second group, by Gruss et al.~\cite{pref_side_channel_attacks_16}, propose introducing privilege checks on prefetch instructions. This would cause a segmentation fault when there is an attempt to prefetch kernel data. This 
prevents access violation in the \emph{interact} step and affect all SW- and HW-based attacks that 
exploit the lack of permission checks~\cite{augury22,Pandora,chen2021leaking,pref_side_channel_attacks_16, amdAttacksSpaceTime}.

\begin{scriptsize}
\begin{table}[b!]
  \small
  \scalebox{0.81}{
  \renewcommand{\arraystretch}{1.16}
  \setlength\tabcolsep{1.5pt} 
  \centering
  \begin{tabular}{|l|c|c|c|c|c|c|c|c|c|}
    \hline
    \textbf{Defense} &\textbf{Res.}  &\multicolumn{4}{c|}{\textbf{Attack step}}&\textbf{Threat}&\textbf{P}\\
     &\textbf{} & \textbf{$A_S$} & \textbf{$A_I$} & \textbf{$A_T$} & \textbf{$A_R$}&\textbf{model} & \\ 
    \hline
   \hline
   disable~\cite{unvelining,amdAttacksSpaceTime,fetchingTale,chen2021leaking} & $R_{Pref.}$ & D/S/I & D/S/A/I & - & D/S/A/I & 1,2,3 & $\CIRCLE$ \\
   privilege checks~\cite{pref_side_channel_attacks_16} & $R_{Pref.}$ & - & A & - & A & 1,2,3 & $\LEFTcircle$ \\
   kernel/user isol.~\cite{KAISER-17} & $R_{TLB}$  & - & S/A/I & - & - & 2 & $\LEFTcircle$\\
   flush~\cite{chen2021leaking,fetchingTale} & $R_{Pref.}$ & I & I & - & I & 2 & $\LEFTcircle$ \\
   replicate~\cite{chen2021leaking,fetchingTale} & $R_{Pref.}$ & S/I & S/I & - & S/I & 2 & $\LEFTcircle$ \\
   const. time~\cite{unvelining,fastertablelookups,constant_time06,practicalMitig} & - & - & D/I & - & - & 1,2,3 & $\LEFTcircle$\\
   \emph{SC} defenses & $R_{SC}$  & D/S/I & - & - & D/S/I  & 1,2,3 & $\LEFTcircle$\\
 \hline
  \end{tabular}
  }
  \caption{Defenses for attacks using the prefetcher.} 
  \label{table:prefeching}
\end{table}
\end{scriptsize}

Another strategy is to provide stronger isolation between kernel and user-space to protect against attacks that leverage the lack of permission checks
~\cite{KAISER-17}. This approach has been adopted in both Linux~\cite{KAISERlinux} and Windows~\cite{KAISERwindows}. This would provide protection against 
SW-based attacks
using the prefetch instruction~\cite{pref_side_channel_attacks_16, amdAttacksSpaceTime} and against HW-based attacks~\cite{augury22,Pandora,chen2021leaking}. This approach restricts sharing and information flow through the page tables (and TLB), in the \emph{interact} step of the attack.

The next group of defenses target attacks that exploit the state of HW-based prefetchers (prefetch tables). Specifically, these defenses replicate and flush prefetcher state at context switches~\cite{chen2021leaking,fetchingTale}. This ensures that state is no longer shared across context switches 
which restricts sharing and information flow. This 
only affects the HW-based attacks which rely on \emph{MS} in the prefetcher. Furthermore, the threat model targeted is limited to \emph{SameThread} since the technique does not affect concurrently executing SMT threads sharing prefetcher state. 



Another strategy is to change the SW implementation to ensure that any prefetch activity is not dependent on any secret. This would affect the attacks relying on data-dependent execution paths and observing prefetch patterns~\cite{unvelining,pref_pref2022,chen2021leaking}. 
This can be achieved using constant-time programming practices~\cite{unvelining,constant_time05}, for example rewriting table based look-ups to be immune to prefetches~\cite{unvelining, fastertablelookups}. This strategy affects determinism and information flow and can theoretically  protect against attacks that exploit prefetch patterns. However, it is challenging to implement this broadly in practice. 

Lastly, it should be noted that some of the attacks rely on the shared state of the \emph{SC}~\cite{unvelining,chen2021leaking,pref_pref2022,augury22,pref_side_channel_attacks_16}. The defenses proposed for the \emph{SC} could be used to defend against these attacks as well. 

\subsubsection{Branch prediction}
\label{def_branch_predictor}

\begin{scriptsize}
\begin{table}[t!]
  \small
  \scalebox{0.81}{
  \renewcommand{\arraystretch}{1.16}
  \setlength\tabcolsep{1.5pt} 
 \centering
  \begin{tabular}{|l|c|c|c|c|c|c|c|c|c|}
    \hline
    \textbf{Defense} &\textbf{Resource}  &\multicolumn{4}{c|}{\textbf{Attack step}}&\textbf{Threat}&\textbf{P}\\
     &\textbf{} & \textbf{$A_S$} & \textbf{$A_I$} & \textbf{$A_T$} & \textbf{$A_R$}&\textbf{model} & \\ 
    \hline
   \hline
   if-conversion\cite{if_conversion_01,BranchScope18} & - & D/S/I & D/S/I & - & D/S/I  & 2,3 & $\CIRCLE$ \\
   enc.:~\cite{lightweightIsolation,twoLevelEncryption} & $R_{BTB}$/$R_{PHT}$ & D/I & D/I & - & D/I & 2,3 & $\LEFTcircle$ \\
   \hspace{7mm}~\cite{BR_exynos,BP_defenceSW16,HyBP_22} & $R_{BTB}$/$R_{PHT}$ & D/I & D/I & - & D/I & 2,3 & $\LEFTcircle$ \\
   flush~\cite{BRB,HyBP_22} & $R_{BTB}$/$R_{PHT}$/$R_{BHB}$ & I & I  & - & I & 2 & $\CIRCLE$ \\
   part.~\cite{BRB,HyBP_22} & $R_{BTB}$/$R_{PHT}$  & S/I & S/I & - & S/I & 2 & $\LEFTcircle$\\
   rand.~\cite{Zhao2021,mitigatingBranchShadowing} & $R_{BTB}$/$R_{PHT}$  & D & D & - & D & 2,3 & $\LEFTcircle$\\ 
 \hline
  \end{tabular}
  }
  \caption{Defenses for attacks using the branch predictor.} 
  \label{table:branch_prediction}
  \vspace{-1.5em}
\end{table}
\end{scriptsize}

The defenses against branch prediction based attacks can be grouped into five categories, see Table \ref{table:branch_prediction}.
The first 
group relies on SW-based techniques, as if-conversion, where the compiler restructures code to avoid conditional branches and use predication instead \cite{if_conversion_01}. This 
restricts determinism, sharing and information flow since branching is avoided and is akin to disabling the direction prediction. 
However, 
the applicability 
to real-world code with complex control flow is limited ~\cite{BranchScope18}. Furthermore, highly predictable branches have been shown to perform poorly when if-converted~\cite{if_conversion_01}.

The next category of defenses 
use randomization to thwart attacks. 
Specifically, encryption of the \emph{BTB}/\emph{BTB} have been proposed~\cite{lightweightIsolation,twoLevelEncryption,BR_exynos,BP_defenceSW16} to prevent the adversary from easily manipulating the branch prediction logic. Encryption restricts determinism and information flow, in all the steps of the attack. Determinism is affected in the \emph{setup} step because 
the target is encrypted which makes manipulating collisions difficult. Information flow is also hindered since a process can only access correct entries in the presence of a valid key. The limitation of these encryption-based solutions is that they cannot guarantee protection against brute-force approaches.


Defenses in the next category flush the state of the branch predictor i.e. \emph{BTB}/\emph{BTB}/\emph{BHB}, on context switches ~\cite{BRB,HyBP_22}. This would affect information flow in the context of the \emph{SameThread} threat model. However, the performance overhead is usually high ~\cite{BRB} and the effectiveness is restricted to the \emph{SameThread} model.  

Partitioning has been shown to thwart attacks on the branch predictor~\cite{BranchScope18,HyBP_22}. Partitioning affects sharing and information flow, in all the steps of the attack, since the state of the \emph{BTB}/\emph{BTB} is isolated.  HyBP~\cite{HyBP_22} combines  isolation and encryption, and selectively replicates parts of the predictor state, while using encryption for the larger tables. The focus of these 
proposals is to 
use partitioning/replication to avoid the high performance cost of flushing the entire branch prediction state upon a context switch. 
Replicating the entire prediction state among SMT threads, although a possibility, is prohibitively expensive.


The last category makes the state transition of the predictor probabilistic, by affecting the saturating counters, as proposed by Zhao et al.\cite{Zhao2021}. This defense restricts determinism in all the steps of the attack. However, the protection offered by the technique is limited since an adversary, through repeated measurements, 
can eventually infer the secret from the state of the branch predictor.

\subsubsection{Computational simplification}
\begin{scriptsize}
\begin{table}[t!]
  \centering
  \small
  \scalebox{0.81}{
  \renewcommand{\arraystretch}{1.16}
  \setlength\tabcolsep{1.5pt} 
  \begin{tabular}{|l|c|c|c|c|c|c|c|c|c|}
    \hline
    \textbf{Defense} &\textbf{Resource}  &\multicolumn{4}{c|}{\textbf{Attack step}}&\textbf{Threat}&\textbf{P}\\
     &\textbf{} & \textbf{$A_S$} & \textbf{$A_I$} & \textbf{$A_T$} & \textbf{$A_R$}&\textbf{model} & \\ 
    \hline
   \hline
   disable~\cite{early_termination_10,practicalMitig} & $R_{MUL}$/$R_{FPU}$ & - & D & - & - & 2,3 & $\CIRCLE$ \\
   const. time:~\cite{practicalMitig,early_termination_10,const_timeFP} & $R_{MUL}$/$R_{FPU}$ & - & D & - & - & 2,3 & $\LEFTcircle$\\
  \hspace{3mm}~\cite{subnornalFP,constant_time05,constant_time06,escort} & $R_{MUL}$/$R_{FPU}$ & - & D & - & - & 2,3 & $\LEFTcircle$ \\
 \hline
  \end{tabular}}
  \vspace{0.01mm}
  \caption{Defenses for attacks using comp. 
  simplification.} 
  \label{table:comp_simp}
  \vspace{-1.5em}
\end{table}
\end{scriptsize}

Two broad strategies have been proposed for protecting against attacks using computational simplification, 
see Table \ref{table:comp_simp}. One 
strategy is to selectively disable the optimization for parts of the program which accesses sensitive information~\cite{early_termination_10,practicalMitig}. Disabling restricts determinism in the \emph{interact} step.

The other strategy is to change the implementation to avoid 
any data-dependent timing variations, even 
for computational simplification. This strategy 
targets 
determinism in the \emph{interact} step. One way to achieve data-independent implementation is to use constant-time programming practices~\cite{subnornalFP,practicalMitig,constant_time05,constant_time06}. 
In~\cite{subnornalFP} a FP library (LibFTFP) is shown, 
providing 
a fixed-point data type with all library operations executing in constant time. 

\textcolor{black}{Lastly, in the case of the browser-based attack~\cite{subnornalFP} the SW-construct which enables the sharing and information flow, can be disabled i.e. cross-origin SVG-filters~\cite{Kohlbrenner17}}.



\vspace{-.2em}
\subsection{Defenses against Transient Attacks}
\vspace{-.2em}

\begin{scriptsize}
\begin{table}[b!]
\vspace{-1.5em}
  \small
  \scalebox{0.81}{
  \renewcommand{\arraystretch}{1.16}
  \setlength\tabcolsep{1.5pt} 
 \centering
  \begin{tabular}{|l|c|c|c|c|c|c|c|c|c|c|}
    \hline
     \textbf{Defense} &\textbf{Resource}  &\multicolumn{4}{c|}{\textbf{Attack step}}&\textbf{Threat}&\textbf{P}\\
     &\textbf{} & \textbf{$A_S$} & \textbf{$A_I$} & \textbf{$A_T$} & \textbf{$A_R$}&\textbf{model} & \\ 
    \hline
   \hline
    local: BPU  & $R_{BTB}$/$R_{PHT}$ & D/S/I & - & - & -  & 2,3 & $\LEFTcircle$ \\ 
    \hspace{6mm} \emph{SC} & $R_{SC}$ & D/S/I & - & - & D/S/I  & 1,2 & $\LEFTcircle$ \\ 
    \hline
     disable: IBPB~\cite{intel_sw}, sb~\cite{arm_sw} & $R_{BTB}$/$R_{PHT}$ & D/S/I & - & - & -  & 1,2,3 & $\LEFTcircle$ \\
    \hspace{9mm} SSBD~\cite{ssbd-amd,intel_sw}& $R_{STL}$ & D/S/I & - & - & -  & 1,2,3 & $\LEFTcircle$ \\
    \hspace{9mm} SMT~\cite{googledisablesmt,openbsddisablesmt,redhatdisablesmt} & - & D/S/I & D/S/I & - & D/S/I & 3 & $\CIRCLE$ \\
    \hline
    restrict: IBRS,STIBP~\cite{amd_sw,intel_sw} & $R_{BTB}$/$R_{PHT}$ & D/S/I & - & - & -  & 1,2,3 & $\LEFTcircle$ \\
    \hspace{9mm}retpoline~\cite{retpoline,amit2019jumpswitches} & $R_{RSB}$  & I & - & - & - & 1,2,3 & $\LEFTcircle$\\
    \hspace{9mm}randpoline~\cite{randpoline_2021} & $R_{RSB}$  & D & - & - & - & 1,2,3 & $\LEFTcircle$\\
     \hspace{9mm}\emph{lfence}, serial.~\cite{intel_sw} & - & - & D/I & - & - & 1,2,3 & $\LEFTcircle$ \\
     fence:~\cite{CSF_19,vassena2020automatically,koruyeh2020speccfi} & $R_{\mu OP\_Q}$ & - & D/I & - & - & 1,2,3 & $\LEFTcircle$ \\
      \hspace{10mm} ~\cite{shen2018restricting,SLH_18,oleksenko2018you} & - & - & D/I & - & - & 1,2,3 & $\LEFTcircle$ \\
     delay:~\cite{NDA_19,SpecShield_19,Schwarz2020ConTExTAG,SpectreGuard_19} & $R_{ROB}$, $R_{reg.}$ & - & D/I/A & - & - & 1,2,3 & $\LEFTcircle$ \\
     \hspace{9mm} ~\cite{STT_19,model_checking19} & $R_{ROB}$, $R_{reg.}$ & - & D/I/A & - & - & 1,2,3 & $\LEFTcircle$ \\
     \hspace{9mm} ~\cite{CondSpec_19,DOLMAS_21,InvSpecDef_19} & $R_{ROB}$, $R_{reg.}$ & - & - & D/I & - & 1,2,3 & $\LEFTcircle$ \\
    rollback~\cite{CleanupSpec} & $R_{SC}$ & - & - & D/I & D/I & 1,2 & $\LEFTcircle$ \\
    \hline
    isolate:~\cite{InvisiSpec_18,SafeSpec_19,MounTrap,GhostMinion_21} & $R_{buf.}$/$R_{L0}$ & - & - & D/S/I & - & 2 & $\LEFTcircle$\\ 
    \hline
  \end{tabular}
  }
  \caption{Classification of defenses for transient attacks.} 
  \label{table:speculative_exec}
   \vspace{-.5em}
\end{table}
\end{scriptsize}


We describe defenses for transient execution attacks, where we first discuss defenses for speculation-based attacks, and then defenses for exception-based attacks.

\subsubsection{Defenses against Speculation-based Attacks}
\label{def_spectre}
The defenses against speculation-based attacks can be broadly grouped into four high-level categories based on the defense strategy: localized defenses, disabling defenses, restriction defenses and isolation defenses, \textcolor{black}{see Table \ref{table:speculative_exec}}.

Localized defenses leverage the defenses available for individual optimizations/resources that interact with speculative execution, such as the BP and/or \emph{SC}. Branch prediction can be targeted in the \emph{setup} step to stop the adversary from being able to (mis)train the predictor, see Section \ref{def_branch_predictor}. The defense is only applicable for the attacks which uses the specific predictor resource. Another approach is to target the side-channel that permits information flow across protection domain through transient execution. In most attacks the \emph{SC} is used since it can offer the highest bandwidth, which makes the defenses 
in Section \ref{def_cache} applicable. 
However, studies have demonstrated that a large variety of microarchitectural resources can be used, such as execution units, ports, \emph{PHT}, \emph{MSHR} etc. This is 
a challenge because multiple resources may need to be protected, 
since they can all act as potential side-channels. By defending the resource which enables the easily accessible (
high bandwidth and low noise) side-channels, the attack bandwidth can be reduced. In addition, limiting the threat model (for instance to \emph{CrossCore}) can restrict the number of resources which can be used as 
side-channels.

The second category involves disabling the optimization. This approach has been proposed in specific scenarios where other defenses are not applicable. Support for disabling indirect branch predictions using a barrier, Indirect Branch Predictor Barrier (IBPB)~\cite{amd_sw,intel_sw}, has been adopted in commodity HW. Likewise, to thwart Spectre v4 attack, the STL mechanism can also be disabled using Speculative Store Bypass Disable (SSBD) microcode updates from Intel and AMD~\cite{ssbd-amd,intel_sw}. These techniques affect the root causes determinism, sharing and information flow in the predictor, in all the steps where it is used. However, the performance cost is potentially high since the predictions are restricted. Another defensive measure is to limit the threat model by disabling SMT
~\cite{googledisablesmt,openbsddisablesmt,redhatdisablesmt}. 
However, disabling SMT comes at a potentially high performance cost.

The third category is restriction-based defences. Here, the high-level idea is to restrict the speculative execution, selectively, to avoid the attacks. Speculation restriction can be performed in the different steps of the attack and using different mechanisms in HW and/or SW. One HW mechanism, adopted in commodity HW to avoid (mis)training, is Indirect Branch Restricted Speculation (IBRS) ~\cite{amd_sw,intel_sw}, which affects the \emph{setup} step. Using IBRS restricts the training of indirect targets inside an enclave. Likewise, using Thread Indirect Branch Prediction (STIBP) restricts the use of prediction entries trained in another SMT thread. Speculative execution have also been restricted using micro-code updates~\cite{intelMicroCodeUpdates}. SW-based defenses~\cite{retpoline,randpoline_2021,amit2019jumpswitches} also aim to restrict  speculation by avoiding the state transition \emph{MS\textsubscript{I}}$\rightarrow$\emph{MS\textsubscript{P}}, and preventing the speculation triggered by the the branch prediction. Here, the (exploitable) indirect branch is replaced by a different retpoline sequence. This will cause the  misspeculated code to execute a controlled loop sequence until speculation has been resolved. To lower the performance cost of the technique both a probabilistic variant, randpoline~\cite{randpoline_2021}, and a HW variant~\cite{amit2019jumpswitches} have been proposed. The root cause exploited is information flow. A recent paper, RETBLEED~\cite{RETBLEED_22}, have shown that the retpoline strategy only provides partial protection, since it can be circumvented using manipulation of return instructions.%

Another mechanism to restrict speculation is to introduce fences to limit transient execution. Many existing attack mitigations use the serializing \emph{lfence} instruction before sensitive parts of the code. In order to improve the usability and performance cost, defenses have been proposed which selectively and automatically choose when to use fences, either in SW~\cite{CSF_19,shen2018restricting,vassena2020automatically,koruyeh2020speccfi} or in HW~\cite{CSF_19}. For example Shen et al.~\cite{shen2018restricting} split code into small blocks and insert fences between the entry point and a potentially leaking memory access to defend against Spectre attacks. \textcolor{black}{The root causes affected are determinism and information flow, depending on the solution.} Another example is Context-Sensitive Fencing (CSF)~\cite{CSF_19} which uses customized decoding from instructions to micro-ops to insert fences after a conditional branch instruction and before a subsequent load instruction. The root causes affected by fences are determinism and information flow, in the \emph{interact} step. Another way similar to fences is Speculative Load Hardening (SLH)~\cite{SLH_18} a compiler-level technique where the idea is to introduce a data dependency on the condition, in order to guarantee that the control flow is valid. The technique is supported in LLVM and GCC~\cite{llvm_gcc}. Oleksenko et al.~\cite{oleksenko2018you} restrict speculation by introducing a data dependency in order to guarantee that a load will only start if the comparison is in registers or L1 cache. However, the technique is only effective if the load is performed after the comparison.

Another method to restrict speculations is to wait for authorization or until data is no longer transient~\cite{NDA_19,SpectreGuard_19,Schwarz2020ConTExTAG,CondSpec_19,InvSpecDef_19,DOLMAS_21,model_checking19} which affect the execution in the \emph{interact} and/or \emph{transmit} steps. Here, the defenses delay to avoid the access violation which leads to leakage (in \emph{interact}) or stall the load which would update \emph{MS} of \emph{SC} (in \emph{transmit}). For example NDA (Non-speculative Data Access)~\cite{NDA_19} provide different policies for controlling control flow and data propagation in \emph{interact}. The root causes affected are determinism, sharing and information flow. SpectreGuard~\cite{SpectreGuard_19} also affects  the \emph{interact} step and proposes to mark secret data and selectively restrict speculation only for data 
from sensitive pages. These defences affect the root cause determinism, access violation and/or information flow in the \emph{interact} step. CondSpec~\cite{CondSpec_19} affects \emph{transmit} by handling loads differently, a load that hits in the \emph{SC} can read the data and complete its execution while a load that experience a cache miss will be stalled and re-issued later. This affects the root cause access violation and/or information flow. 
CleanUpSpec~\cite{CleanupSpec}, on the contrary, restricts how speculative updates are encoded in the \emph{MS} of the \emph{SC}. Speculative accesses are allowed to progress and make changes to the \emph{SC} but these are removed in case of miss-speculation. This has been shown to be insufficient in certain conditions~\cite{Speculative_interference_attacks_21}. This defense affects the root causes determinism and information flow, in the \emph{transmit} and \emph{receive} step of the attack. 

The performance cost of the restriction-based defenses depends on how restrictive the rule set is, i.e., how much execution differs from the unconstrained speculation scenario. Introducing protections at a later attack step generally leads to more flexibility, since more speculation can be allowed, and comes at a lower performance cost \cite{sok_hw_defences_21}. The trade-off is that more possible side-channels can be used for the state transitions \emph{MS\textsubscript{I}}$\rightarrow$\emph{MS\textsubscript{T}} and \emph{MS\textsubscript{T}}$\rightarrow$\emph{MS\textsubscript{R}}, which makes ensuring full protection challenging.

The last category targets isolation by introducing a shadow structure to hold the speculative \emph{MS} until it is deemed safe~\cite{InvisiSpec_18,SafeSpec_19,MounTrap,GhostMinion_21}. For example, in MounTrap~\cite{MounTrap} an L0 filter cache is used for speculative data, which is only allowed to propagate to the rest of the cache hierarchy after commit. This allows the speculation and potential access violation to occur but not affect \emph{MS} of the \emph{SC}, for example. This restricts information flow from the transient execution to the non-transient execution.

\subsubsection{Defenses against Exception-based Attacks}
\label{def_meltdown}

Exception-based attacks typically exploit implementation oversights in HW.  
The Meltdown attack exploits a race condition between authorization and access~\cite{new_models_20} which enables transient execution to continue with an unauthorized value, leading to access violation. Newer CPUs contain patches whereas existing ones are mostly protected through microcode updates or other workarounds~\cite{EvolutionDefencesTransient20}. For instance, the issue has been addressed on newer Intel microarchitectures~\cite{intel_sw},  Whiskey Lake and onward, by returning zero when accessing privileged memory~\cite{KASLR_FIX_20}. 
In the case of the MDS-attacks the leaks 
is attributed to a use-after-free vulnerability where stale data is read in the internal registers~\cite{ZombieLoad_mds_19}, allowing unintended information flow through shared CPU buffers. The issue is solved by flushing the internal buffers to restrict the information flow. Similarly, to defend against LazyFP, FPU registers are flushed on context switches when changing protection domains with SGX, for hypervisors and for logical cores. In addition, since Linux 4.6 eager FPU switching is used by default~\cite{fpuEager}. This disables the fault since the FPU is always available. 
Foreshadow has been mitigated on Intel CPUs through setting a physical page number field of unmapped page-tables to refer to non-existing physical memory~\cite{foreshadowINTEL,foreshadowNG_18}, thereby restricting access violation. The LVI attacks~\cite{LV1_20} are no longer possible when the corresponding fault or microcode assist is mitigated. 

Another defense strategy is to provide stronger address space isolation~\cite{KAISER-17,LAZARUS_17,EPTI_18,MemoryRanger_18}, which mitigates or limits the possible access violation in the attacks. For example MemoryRanger~\cite{MemoryRanger_18} isolates drivers, kernel and user space into separate address spaces using Extended Page Table (EPT). This defense restricts access violation in the \emph{interact} step, by providing isolation.

\vspace{-.3em}
\subsection{System-level defenses}
\vspace{-.3em}
Broader system-level strategies for providing protection have also been proposed. Leveraging constant-time programming paradigm is one such strategy where the key idea is to rewrite the SW-implementation to avoid timing variability. The challenge with constant-time programming is to provide protection for all types of attacks.
Another strategy is to decrease the accuracy of timing measurement~\cite{fuzzy_time_91,TimeWarp_12,trustedBrowser16,BiggerFish22}. \textcolor{black}{
This strategy affects determinism in the \emph{receive} step,} and can be used as a defense against several attacks on different optimizations. The strategy leads to lower attack bandwidth but has been shown to be ineffective in providing complete protection since many attacks can amplify the timing difference ~\cite{spectre_is_here_to_stay_19} or use other timer mechanisms~\cite{timerInJavascript}. Another strategy is to use formal models~\cite{FormalModel_speculation_19,SPECTECTOR_18,new_models_20,contracts_21,axiomaticContrscts22} in order to 
automatically synthesize vulnerabilities. This strategy could potentially find all exploitable scenarios in the design phase itself. However, identifying all vulnerabilities complicate the model building process and current proposals therefore only target a limited set. 
Finally, leveraging programming language based security~\cite{kozen1999language} is another option wherein defenses are co-designed  leveraging programming language annotations to mark sensitive parts of the programs, along with suitable HW primitives. However, a recent analysis performed by Naseredini et al.~\cite{ProgrammingLanguagesSpectreMitigations21} found that most programming languages and their execution environments does not have support for Spectre mitigations. This shows the challenge of relying on programming languages and execution environments to provide complete protection.

\vspace{-.3em}
\section{Discussion}
\vspace{-.3em}
\label{discussion}

\textit{Commonalities:} We have shown that the root causes that enable the attacks are common, across a wide range of microarchitectural optimizations. Furthermore, through our analysis of the proposed defenses, we have shown that these target one or more root cause, across the different steps of an attack. 
There are commonalities even in the defense strategies used to protect against attacks on these diverse microarchitectural optimizations. Some of those include disabling the optimization, to restrict all the root causes; isolating the state related to the optimization, to restrict sharing; applying randomization and/or restriction, to limit information flow and introducing permission checks, to limit resources from, exposing/accessing state outside of the intended domain.

Using these common strategies together with the root causes, enable us to envision new defenses for vulnerable microarchitectural optimizations. 
We apply these common strategies to defend against potential attacks exploiting value prediction (Section \ref{attack_value_prediction}). 
Possible defenses against those attacks could involve flushing the table at context switches, isolating and/or partitioning the table 
to avoid conflict and reuse-based attacks, introducing randomization to limit information flow. Another option could involve introducing non-determinism in the value prediction mechanisms. Lastly, the simplest mitigation would be to provide mechanisms to selectively disable the optimization when running sensitive parts of the program.

\textit{Observations:}
Firstly, the ease of exploiting a microarchitectural optimization and  the severity of the leak vary widely. There exist limiting factors which are not easy to quantify, such as the availability and capabilities of gadgets in the victim code~\cite{SysEvauationTransient19}. The link between attack bandwidth and effectiveness in practical scenarios is also difficult to quantify.


Secondly, we observe that, in several cases, 
vulnerabilities are not due to the fundamental behaviour of the microarchitectural optimization but are rather a result of design and/or implementation. This points in the direction of promoting a deeper understanding of the root causes of vulnerabilities and the potential defense strategies in the design and implementation phases, which we hope can be assisted by our framework, rather than as an afterthought. We believe that microarchitectural optimizations can still be a promising avenue to provide 
performance scalability in future technology nodes without having to compromise on security. 

\textit{Future work:} 
Our focus, in this paper, has been on vulnerabilities in microarchitectural optimizations that target performance. We have not exhaustively covered all microarchitectural optimizations and/or resources, for example the NoC and DRAM. However, we expect our framework to be easily extended to cover attacks and defenses on other optimizations and resources. We have not focused on SW vulnerabilities, such as buffer overflows, which poses considerable security risks. Furthermore, we have not investigated microarchitectural optimizations for security, such as Intel SGX, nor power-based side channel attacks \textcolor{black}{nor performance degradation attacks~\cite{Grunwald2002,DoS_CACHE,AmplifyingByPerfDegradation,HyperDegrade}}. Investigating and/or extending the root cause framework to include optimizations for security and considering power-side channels is left for future work.

\vspace{-.3em}
\section{Conclusions}
\vspace{-.3em}
\label{conclusions}
We identify four root causes that enable timing-based side-channel attacks on an extensive set of microarchitectural optimizations. We provide a framework and systematize both transient and non-transient attacks and defenses, highlighting the similarities and differences. Based on our analysis we discuss potential attacks and defenses for vulnerable optimizations. We believe our framework can assist computer architects in understanding the landscape of attacks and defenses, and to provide guidance in designing secure microarchitectural optimizations. 


\bibliographystyle{plain}
\bibliography{refs}

\end{document}